\documentclass[aps,prmaterials,twocolumn,longbibliography,nobibnotes,showkeys]{revtex4-2}

\usepackage{graphicx}% Include figure files
\usepackage{dcolumn}% Align table columns on decimal point
\usepackage{bm}% bold math
\usepackage[version=3]{mhchem}
\usepackage{color}

\begin{document}
\title{Rare-earth defects in GaN: A systematic investigation of the lanthanide series}
\author{Khang Hoang}
\email{khang.hoang@ndsu.edu}
\affiliation{Center for Computationally Assisted Science and Technology \& Department of Physics, North Dakota State University, Fargo, North Dakota 58108, United States}

\date{\today}

\begin{abstract}

Rare-earth (RE) doped GaN is of interest for optoelectronics and spintronics and potentially for quantum applications. A fundamental understanding of the interaction between RE dopants and the semiconductor host is key to realizing the material's full potential. This work reports an investigation of lanthanide ($Ln$) defects in GaN using hybrid density-functional defect calculations. We find that all the $Ln$ dopants incorporated at the Ga lattice site, $Ln_{\rm Ga}$ ($Ln$ = La--Lu), are stable as trivalent ions, but Eu and Yb can also be stabilized as divalent and Ce, Pr, and Tb as tetravalent. The location of $Ln$-related defect levels and the $Ln$ $4f$ states in the energy spectrum of the host material is determined from first principles. We elucidate the interplay between defect formation and electronic structure, including the $Ln$--N interaction, and the effect of doping on the local lattice environment. Optical properties are investigated by considering possible defect-to-band and band-to-defect transitions involving $Ln_{\rm Ga}$ defects with in-gap energy levels, including broad ``charge-transfer'' transitions. These defects can also act as carrier traps and mediate energy transfer from the host into the $4f$-electron core of the $Ln$ ion which leads to sharp intra-$f$ luminescence.  

\end{abstract}

\pacs{}
%\keywords{}

\maketitle


\section{Introduction}\label{sec;intro}

Rare-earth (RE) doped semiconductors have long been of interest for optoelectronics and spintronics \cite{ODonnell2010Book}. In the RE impurities, the $4f$-electron core is well shielded by the outer $5s^2$ and $5p^6$ electron shells, resulting in very sharp intra-$f$ optical transitions at wavelengths from infrared to ultraviolet. GaN doped with Pr, Eu, Er, or Tm, for example, emits light in a few narrow bands in the visible spectrum \cite{StecklMRS1999}. In addition to the $4f$--$4f$ transitions, ``charge-transfer'' and  $5d$--$4f$ transitions can also occur in RE-doped luminescent materials \cite{Blasse1994Book}. GaN has also been identified as a promising host material for defect-based qubits \cite{Weber2010PNAS,Gordon2013MRSBull}, mainly due to its wide band gap and weak spin-orbit coupling. Defects suitable for quantum applications are not limited to native point defects and non-RE impurities but can also be RE impurities which, in addition to the sharp optical transitions, have excellent spin coherence properties. Although there has been intensive research on RE-doped complex oxide insulators for quantum computing and optical quantum memories \cite{Thiel2011JL,Kunkel2018ZAAC,Zhong2019NP}, prospects of RE-doped GaN for quantum applications were discussed only recently \cite{Mitchell2021NP}. Whether a RE dopant is being harnessed for traditional optical applications or novel quantum technologies, having a fundamental understanding of the interaction between the dopant and the host material is key to realizing its potential.

The location of RE-related defect levels in the energy spectrum of the semiconductor host is important information to understand and predict the material's properties. Yet, in RE-doped GaN, direct information from experiments has been limited. McHale et al.~\cite{McHale2011EPJAP} reported that the occupied Gd, Er, and Yb $4f$ states are deep in the valence band in GaN thin films. There were reports of a broad ``charge-transfer'' excitation or absorption band associated with the Eu defect in Eu-doped GaN \cite{Morishima1999PSS,Tanaka2003PSS,Nyein2003APL,Higuchi2010PRB,Li2002JCG,Sawahata2005STAM}, which can provide the location of the Eu$^{3+/2+}$ level. On the basis of an semi-empirical model, Dorenbos and van der Kolk proposed a scheme with the location of all lanthanide impurity levels in GaN \cite{Dorenbos2006APL}. Although such a scheme has been fairly successful in explaining certain properties of the material, a more rigorous methodology and, more importantly, a deeper understanding of RE-related defect structure and energetics are needed if further advances are to be made in understanding and designing RE-doped functional materials. First-principles calculations based on density functional theory (DFT) can be extremely useful in supporting such progress.

Computational studies of RE-doped semiconductors and insulators have been very challenging, however. This is due to the requirement to properly describe both the host states, including the band gap, and the impurity states, including the highly localized RE $4f$ states, in the doped materials. Standard DFT calculations within the local-density (LDA) or generalized gradient (GGA) approximation \cite{LDA1980,PW91} are not suitable for systems with partially filled $4f$ orbitals. Even the Hubbard-corrected DFT$+$$U$ method \cite{anisimov1991} fails to satisfactorily describe their basic defect physics \cite{Hoang2021PRM}, mainly due to the fact that the Hubbard $U$ term is applied on the RE $4f$ states only and all other orbitals are left uncorrected. Only recently, a hybrid DFT/Hartree-Fock approach \cite{heyd:8207} has been applied successfully to provide reliable results for Eu- and Er-related defects in GaN \cite{Hoang2021PRM,Hoang2015RRL,Hoang2016RRL} (See the cited references for a thorough comparison of the results obtained in calculations using hybrid functional vs.~other DFT-based methods. Previously, the hybrid functional approach was also reported to provide a ``balanced description'' of the electronic properties of bulk materials such as RE oxides \cite{DaSilva2007PRB}). Many other RE defects with potentially interesting and useful properties remain to be explored.   

We herein present an investigation of RE defects in wurtzite GaN using hybrid density-functional defect calculations where all orbitals in the material are treated on equal footing. Specific calculations are carried out for substitutional lanthanide ($Ln$) impurities at the Ga lattice site, i.e., $Ln_{\rm Ga}$ ($Ln$ = La--Lu). The interstitial $Ln$ defects are not considered here because they are expected to have high formation energies and thus unlikely to form. On the basis of our results, we discuss the atomic and electronic structure, energetics, and optical properties of $Ln_{\rm Ga}$. Comparison with experiments and previous computational work will be included where appropriate. 

\section{Methodology}\label{sec;method} 

Point defects are modeled using a supercell approach in which a defect is included in a periodically repeated finite volume of the host material. The formation energy of a general defect X in charge state $q$ (with respect to the host lattice) is defined as \cite{Freysoldt2014RMP}
\begin{align}\label{eq:eform}
E^f({\mathrm{X}}^q)&=&E_{\mathrm{tot}}({\mathrm{X}}^q)-E_{\mathrm{tot}}({\mathrm{bulk}}) -\sum_{i}{n_i\mu_i^*} \\ %
\nonumber &&+~q(E_{\mathrm{v}}+\mu_{e})+ \Delta^q ,
\end{align}
where $E_{\mathrm{tot}}(\mathrm{X}^{q})$ and $E_{\mathrm{tot}}(\mathrm{bulk})$ are the total energies of the defect-containing and bulk supercells. $n_{i}$ is the number of atoms of species $i$ that have been added ($n_{i}>0$) or removed ($n_{i}<0$) to form the defect. $\mu_{i}^*$ is the atomic chemical potential, representing the energy of the reservoir with which atoms are being exchanged, and referenced to the total energy per atom of $i$ in its elemental phase at 0 K; e.g., $\mu_{\rm Ga}^* = E_{\rm tot}({\rm Ga}) + \mu_{\rm Ga}$, with $E_{\rm tot}({\rm Ga})$ being the total energy per atom of metallic Ga, and $\mu_{\rm Ga} \le 0$ eV. $\mu_{e}$ is the chemical potential of electrons, i.e., the Fermi level, representing the energy of the electron reservoir, and referenced to the valence-band maximum (VBM) in the bulk ($E_{\mathrm{v}}$). $\Delta^q$ is the correction term to align the electrostatic potentials of the bulk and defect-containing supercells and to account for finite-size effects on the total energies of charged defects \cite{Freysoldt,Freysoldt11}.

The {\it thermodynamic} transition level between charge states $q$ and $q'$ of a defect, $\epsilon(q/q')$, is defined as the Fermi-level position at which the formation energy of the defect in charge state $q$ is equal to that in state $q'$ \cite{Freysoldt2014RMP}, i.e.,
\begin{equation}\label{eq;tl}
\epsilon(q/q') = \frac{E^f(X^{q}; \mu_e=0)-E^f(X^{q'}; \mu_e=0)}{q' - q},
\end{equation}
where $E^f(X^{q}; \mu_e=0)$ is the formation energy of the defect X in charge state $q$ when the Fermi level is at the VBM ($\mu_e=0$). This $\epsilon(q/q')$ level [also referred to as the $(q/q')$ level], corresponding to a {\it defect energy level} (or, simply, {\it defect level}), would be observed in experiments where the defect in the final charge state $q'$ fully relaxes to its equilibrium configuration after the transition. 

From the defect formation energy, one can also calculate the {\it optical} transition level $E_{\rm opt}^{q/q'}$, which is employed to characterize defect-to-band and band-to-defect optical transitions. The level is defined similarly to $\epsilon(q/q')$ but with the total energy of the final state $q'$ calculated using the lattice configuration of the initial state $q$ \cite{Freysoldt2014RMP}. 

The total-energy electronic structure calculations are based on DFT with the Heyd-Scuseria-Ernzerhof (HSE) functional \cite{heyd:8207}, the projector augmented wave method \cite{PAW2}, and a plane-wave basis set, as implemented in the Vienna {\it Ab Initio} Simulation Package (\textsc{vasp}) \cite{VASP2}. We use the standard PAW potentials in the \textsc{vasp} ({\it potpaw\_PBE.54}) database which treat the $Ln$ $4f$ electrons explicitly as valence electrons.  Like in our previous work \cite{Hoang2016RRL,Hoang2021PRM}, the Hartree-Fock mixing parameter ($\alpha$) is set to 0.31 and the screening length to the default value of 10 {\AA} to match the host's experimental band gap ($\sim$3.5 eV). RE defects in the GaN host are simulated using a 96-atom supercell. In such a supercell model for the substitutional $Ln$ impurity ($Ln_{\rm Ga}$), one Ga atom is substituted with $Ln$ and thus the chemical composition is $Ln$Ga$_{47}$N$_{48}$; i.e., the doping concentration is $\sim$2\%. In the defect calculations, the lattice parameters are fixed to the calculated bulk values but all the internal coordinates are relaxed. All structural relaxations are performed with HSE and the force threshold is chosen to be 0.02 eV/{\AA}. The plane-wave basis-set cutoff is set to 400 eV and spin polarization is included. Spin-orbit coupling (SOC) is not included; it was previously shown that SOC had negligible effects on the defect transition levels \cite{Hoang2021PRM}. We employ a 2$\times$2$\times$2 Monkhorst-Pack $k$-point mesh for the integrations over the Brillouin zone, except in the calculations to obtain the electronic density of states where a denser, $\Gamma$-centered 3$\times$3$\times$3 $k$-point mesh is used.

\begin{table}%[b]
\caption{Formation enthalpies of rare-earth mononitrides, calculated at 0 K. All the $Ln$N binaries are assumed to be in the $Fm\overline{3}m$ space group. The standard heats of formation are also included. The unit is in eV per formula unit}\label{tab;enthalpy}
\begin{center}
\begin{ruledtabular}
\begin{tabular}{lcccc}
 &Magnetic order &$\Delta H$ (calc.) &$\Delta H$ (expt.)$^a$ \\
\colrule
LaN &NM & $-2.98$ &$-3.13\pm 0.56$ \\
CeN &FM & $-3.98$ &$-3.39\pm 0.74$ \\
PrN &AF-I & $-3.50$ &$-3.01\pm 0.50$ \\
NdN &FM & $-3.30$ &$-3.10\pm 0.49$ \\
PmN &AF-II & $-3.43$ &$-3.32\pm 0.65$ \\
SmN &AF-I & $-2.79$ &$-3.35\pm 0.16$ \\
EuN &AF-I & $-1.25$ &$-2.02\pm 0.12$ \\
GdN &FM & $-3.42$ &$-3.18\pm 0.22$ \\
TbN &FM & $-3.63$ &$-3.10\pm 0.54$ \\
DyN &AF-I & $-3.83$ &$-3.42\pm 0.53$ \\
HoN &AF-II & $-3.96$ &$-3.53\pm 0.20$ \\
ErN &AF-II & $-4.37$ &$-3.71\pm 0.23$ \\
TmN &AF-I & $-3.84$ &$-3.70\pm 0.53$ \\
YbN &AF-II & $-2.48$ &$-3.74\pm 0.14$ \\
LuN &NM & $-3.86$ &$-3.02\pm 0.54$ \\
\end{tabular}
\end{ruledtabular}
\end{center}
\begin{flushleft}
$^a$Ref.~\citenum{Kordis1977JNM}
\end{flushleft}
\end{table}

\begin{figure*}%[t]%
\vspace{0.2cm}
\includegraphics*[width=\linewidth]{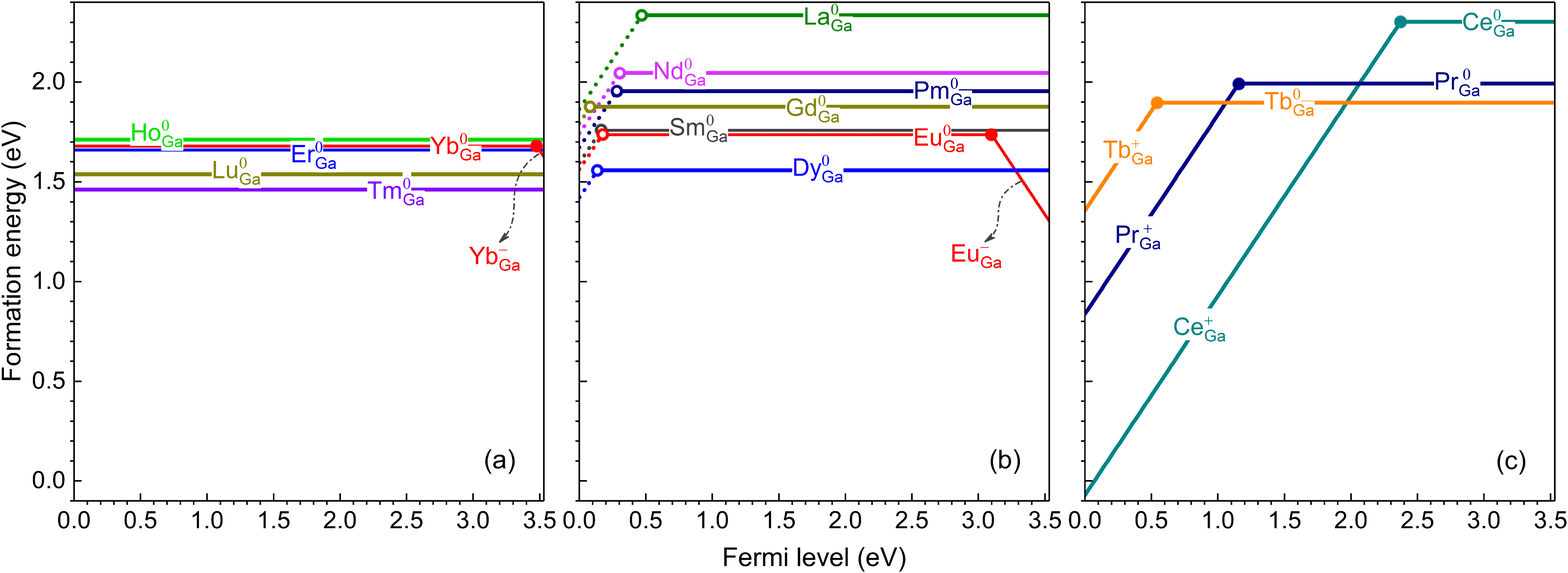}
\caption{Formation energies of $Ln_{\rm Ga}$ in GaN, plotted as a function of the Fermi level from the VBM (at 0 eV) to the conduction-band minimum (CBM, at 3.53 eV): (a) $Ln$ = Ho, Er, Tm, Yb, and Lu, (b) $Ln$ = La, Nd, Pm, Sm, Eu, Gd, and Dy, and (c) $Ln$ = Ce, Pr, and Tb. For each defect, only segments of the formation energy lines corresponding to the lowest-energy charge states are shown. The slope of these segments indicates the charge state $q$: positively (negatively) charged defect configurations have positive (negative) slopes; horizontal segments correspond to neutral defect conﬁgurations. The dotted formation energy lines correspond to defect configurations ($Ln_{\rm Ga}^+$, not indicated in the figure) that consist of an $Ln$ ion at the Ga site and an electron hole localized at a nearby N site. Large dots connecting two segments with different slopes, if present, mark the {\it defect levels} in the host band gap [i.e., the thermodynamic transition level $\epsilon(q/q')$, calculated according to Eq.~(\ref{eq;tl})].}
\label{fig;fe} 
\end{figure*}

The chemical potentials of Ga and N vary over a range determined by the calculated formation enthalpy of GaN: $\mu_{\rm Ga}+\mu_{\rm N} = \Delta H({\rm  GaN}) (-1.23$ eV at 0 K). The extreme Ga-rich and N-rich conditions correspond to $\mu_{\rm Ga} = 0$ eV and $\mu_{\rm N} = 0 $ eV where GaN is assumed to be in equilibrium with metallic Ga and an isolated N$_2$ molecule, respectively. With a given $\mu_{\rm N}$ value, the atomic chemical potential of $Ln$, $\mu_{Ln}$, is obtained by assuming equilibrium with $Ln$N (space group $Fm\overline{3}m$). Table \ref{tab;enthalpy} lists the formation enthalpy of $Ln$N calculated within the HSE functional (with $\alpha = 0.31$); the lowest-energy magnetic structure of $Ln$N is found to be either type-I (AF-I) or type-II (AF-II) antiferromagnetic \cite{Phani1980PRB,Hoang2007PRL}, ferromagnetic (FM), or nonmagnetic (NM). Given the phase equilibrium assumption, the formation energy of $Ln_{\rm Ga}$ is the same for the Ga-rich and N-rich conditions. 

Finally, it should be noted that the thermodynamic and optical transition levels are independent of the choice of the chemical potentials. Also, effects of possible corrections to the total energies beyond the level of theory employed in the current work are expected to be small due to cancellation between different terms in Eq.~(\ref{eq;tl}). 

\section{Results and discussion}\label{sec;results}

We begin by summarizing the basic properties of the host material. In wurtzite GaN, Ga is tetrahedrally coordinated with N atoms: one along the $c$-axis and three in the basal ($ab$) plane. The calculated axial and basal Ga$-$N bond lengths are 1.958 {\AA} and 1.952 {\AA}, respectively, which are consistent with the experimental values (1.956 {\AA} and 1.949 {\AA}) \cite{Schulz1977SSC}. There is thus a small $C_{3v}$ distortion at the Ga lattice site. The calculated band gap is 3.53 eV, a direct gap at the $\Gamma$ point. In the following, we discuss the energetics of the RE defects and how the structural, electronic, and optical properties of the host material is modified by the presence of a RE dopant.

\subsection{Defect energy levels}\label{sec;energetics}

Figure \ref{fig;fe} shows the formation energies of $Ln_{\rm Ga}$ in GaN. We divide the substitutional defects into three groups (A, B, and C) based on characteristics of the defects' energetics near the VBM. As it becomes clearer later, the features near the VBM (which consists predominantly of the N $2p$ states) strongly reflect the $Ln$--N interaction.

\begin{table}%[b]
\caption{Electronically stable $Ln$ ions (and their spin $S$) in the $Ln_{\rm Ga}$ defect and thermodynamic transition levels of $Ln_{\rm Ga}$ in the host band gap (in eV, with respect to the VBM).}\label{tab;re}
\begin{center}
\begin{ruledtabular}
\begin{tabular}{cccccc}
Defect &\multicolumn{2}{c}{$Ln$ ion} & Spin &$\epsilon(+/0)$ & $\epsilon(0/-)$ \\
\colrule
La$_{\rm Ga}$ &La$^{3+}$ & $4f^0$ & $0$ & 0.47 & \\
Ce$_{\rm Ga}$ &Ce$^{4+}$ & $4f^0$ & $0$ & & \\
							&Ce$^{3+}$ & $4f^1$ & $1/2$ & 2.37 & \\
Pr$_{\rm Ga}$ &Pr$^{4+}$ & $4f^1$ & $1/2$ & & \\
							&Pr$^{3+}$ & $4f^2$ & $1$ & 1.16 &\\
Nd$_{\rm Ga}$ &Nd$^{3+}$ & $4f^3$ & $3/2$ & 0.30 & \\
Pm$_{\rm Ga}$ &Pm$^{3+}$ & $4f^4$ & $2$ & 0.28 & \\
Sm$_{\rm Ga}$ &Sm$^{3+}$ & $4f^5$ & $5/2$ & 0.16 & \\
Eu$_{\rm Ga}$ &Eu$^{3+}$ & $4f^6$ & $3$ & 0.21 & \\
							&Eu$^{2+}$ & $4f^7$ & $7/2$ & & 3.10 \\
Gd$_{\rm Ga}$ &Gd$^{3+}$ & $4f^7$ & $7/2$ & 0.08 & \\
Tb$_{\rm Ga}$ &Tb$^{4+}$ & $4f^7$ & $7/2$ & & \\
							&Tb$^{3+}$ & $4f^8$ & $3$ & 0.54 & \\
Dy$_{\rm Ga}$ &Dy$^{3+}$ & $4f^9$ & $5/2$ & 0.14 & \\
Ho$_{\rm Ga}$ &Ho$^{3+}$ & $4f^{10}$ & $2$ & & \\
Er$_{\rm Ga}$ &Er$^{3+}$ & $4f^{11}$ & $3/2$ & & \\
Tm$_{\rm Ga}$ &Tm$^{3+}$ & $4f^{12}$ & $1$ & & \\
Yb$_{\rm Ga}$ &Yb$^{3+}$ & $4f^{13}$ & $1/2$ & & \\
							&Yb$^{2+}$ & $4f^{14}$ & $0$ & & 3.48 \\
Lu$_{\rm Ga}$ &Lu$^{3+}$ & $4f^{14}$ & $0$ & & \\
\end{tabular}
\end{ruledtabular}
\end{center}
\begin{flushleft}
\end{flushleft}
\end{table}

\begin{figure*}%[h]
\centering
\includegraphics[width=\linewidth]{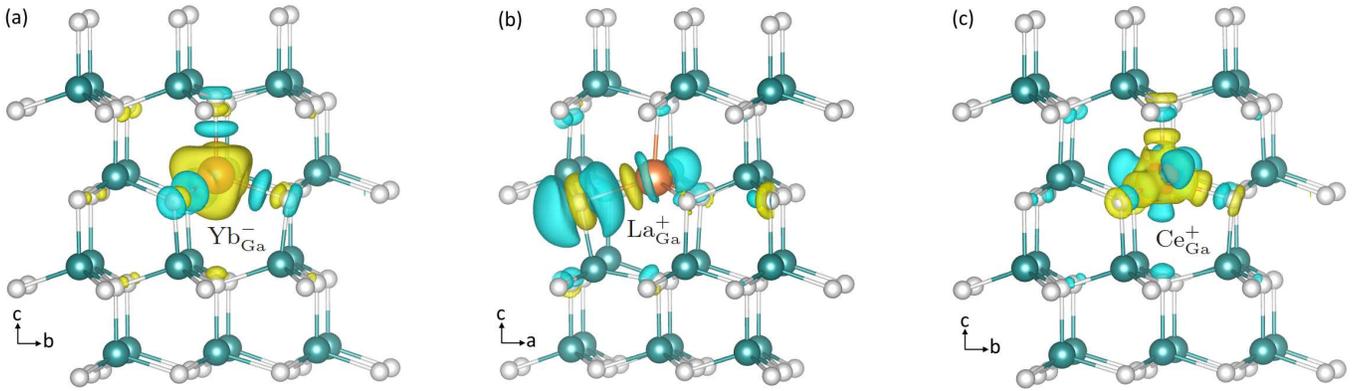}
\caption{Structure of representative $Ln_{\rm Ga}$ configurations: (a) Yb$_{\rm Ga}^-$, (b) La$_{\rm Ga}^+$, and (c) Ce$_{\rm Ga}^+$. The charge density, taken with respect to that of the respective neutral defect configuration but calculated using the lattice environment of the charged one, shows a localized (Yb $4f$) electron (in the case of Yb$_{\rm Ga}^-$), (N $2p$) hole (La$_{\rm Ga}^+$), or (Ce $4f$) hole (Ce$_{\rm Ga}^+$). The isovalue for the charge-density isosurface is set to 0.02 e/{\AA}$^3$. Large (red/green) spheres are $Ln$/Ga and small (gray) spheres are N.}
\label{fig;struct}
\end{figure*}

Group A consists of $Ln$ = Ho, Er, Tm, Yb, and  Lu, i.e., the last five elements in the lanthanide series. In this group, $Ln_{\rm Ga}$, except $Ln=$ Yb, is structurally, electronically, and energetically stable only as $Ln_{\rm Ga}^0$ (i.e., the trivalent $Ln^{3+}$ at the Ga site) and does not have any defect level in the host band gap; see Fig.~\ref{fig;fe}(a). Yb$_{\rm Ga}$ is structurally and electronically stable as Yb$_{\rm Ga}^0$ (i.e., the trivalent Yb$^{3+}$ at the Ga site) and Yb$_{\rm Ga}^-$ (i.e., the divalent Yb$^{2+}$ at the Ga site), and the valence change occurs at the $(0/-)$ level at 0.05 eV below the CBM (i.e., 3.48 eV above the VBM); see Fig.~\ref{fig;fe}(a) and Table \ref{tab;re}. Yb$_{\rm Ga}^0$ is thus energetically more favorable than Yb$_{\rm Ga}^-$ in almost the entire range of the Fermi-level values from the VBM to the CBM, except in a very small range below the CBM in which Yb$_{\rm Ga}^-$ is more favorable. Figure \ref{fig;struct}(a) clearly shows the lattice geometry of and the charge density associated with the localized ($4f$) electron in Yb$_{\rm Ga}^-$, thus confirming the stabilization of Yb$^{2+}$. In general, we determine the valence of a RE ion in a defect configuration by examining the total and local magnetic moments, electron occupation, and local lattice environment. Note that the charge-density behavior of Yb$_{\rm Ga}^-$ is similar to that of the ``atomic-like dopant'' described in Lyons et al.~\cite{Lyons2021JAP}. 

Group B consists of $Ln$ = La, Nd, Pm, Sm, Eu, Gd, and Dy. Each of the $Ln_{\rm Ga}$ defects in this group introduces a defect level, $(+/0)$, just above the VBM; see Fig.~\ref{fig;fe}(b) and the $\epsilon(+/0)$ values explicitly listed in Table \ref{tab;re}. Above the $(+/0)$ level, $Ln_{\rm Ga}$ is energetically more favorable as $Ln_{\rm Ga}^0$ (i.e., $Ln^{3+}$ at the Ga site); below the $(+/0)$ level, $Ln_{\rm Ga}$ is more favorable as $Ln_{\rm Ga}^+$. It is, however, noted that $Ln_{\rm Ga}^+$ here is {\it not} a true charge state of $Ln_{\rm Ga}$, but a defect complex consisting of $Ln_{\rm Ga}^0$ and an electron hole localized on the N atom (hereafter referred to as $h^\ast$, with spin $S=1/2$) that is basally bonded to $Ln$; thus $Ln_{\rm Ga}^+$ = $Ln_{\rm Ga}^0 + h^\ast$. Figure \ref{fig;struct}(b) shows the lattice geometry of La$_{\rm Ga}^+$ and the charge density associated with $h^\ast$. The charge density for Nd$_{\rm Ga}^+$, Pm$_{\rm Ga}^+$, Sm$_{\rm Ga}^+$, Eu$_{\rm Ga}^+$, Gd$_{\rm Ga}^+$, and Dy$_{\rm Ga}^+$ is similar to that for La$_{\rm Ga}^+$. Note that the localized hole state (and hence the $Ln_{\rm Ga}^+$ configuration) is stable even in HSE calculations with smaller mixing parameters (e.g., $\alpha = 0.25$). Given the charge-density characteristic, these $Ln_{\rm Ga}^+$ defects are thus similar to the ``polaronic dopant'' discussed in Ref.~\citenum{Lyons2021JAP}. Eu$_{\rm Ga}$ introduces another defect level, $(0/-)$, at 0.43 eV below the CBM (i.e., 3.10 eV above the VBM), above which Eu$_{\rm Ga}^-$ (i.e., Eu$^{2+}$ at the Ga site) is energetically more favorable than Eu$_{\rm Ga}^0$. The charge density for Eu$_{\rm Ga}^-$ is similar to that for Yb$_{\rm Ga}^-$ [Fig.~\ref{fig;struct}(a)]; and, like Yb$_{\rm Ga}^-$, Eu$_{\rm Ga}^-$ is similar to the ``atomic-like dopant'' \cite{Lyons2021JAP}. 

Group C consists of $Ln$ = Ce, Pr, and Tb. Each of these $Ln_{\rm Ga}$ defects introduces one defect level, $(+/0)$, in the host band gap; see Fig.~\ref{fig;fe}(c) and the $\epsilon(+/0)$ values explicitly listed in Table \ref{tab;re}. Above the $(+/0)$ level, Ce$_{\rm Ga}$, Pr$_{\rm Ga}$, and Tb$_{\rm Ga}$ are energetically more favorable as Ce$_{\rm Ga}^0$, Pr$_{\rm Ga}^0$, and Tb$_{\rm Ga}^0$, i.e., the trivalent $Ln^{3+}$ ion at the Ga site; below the $(+/0)$ level, they are more favorable as Ce$_{\rm Ga}^+$, Pr$_{\rm Ga}^+$, and Tb$_{\rm Ga}^+$ , i.e., the tetravalent $Ln^{4+}$ ion at the Ga site. The $Ln_{\rm Ga}^+$ configuration here is, therefore, a true charge state of $Ln_{\rm Ga}$. The transition from the neutral to positive charge state is thus associated with valence change on the lanthanide ion. Figure \ref{fig;struct}(c) shows the lattice geometry of and the charge density associated with the localized ($4f$) hole in Ce$_{\rm Ga}^+$. The charge density for Pr$_{\rm Ga}^+$ and Tb$_{\rm Ga}^+$ is similar to that for Ce$_{\rm Ga}^+$. The charge-density behavior of Ce$_{\rm Ga}^+$, Pr$_{\rm Ga}^+$, and Tb$_{\rm Ga}^+$ is thus similar to that of the ``atomic-like dopant'' \cite{Lyons2021JAP}.                 

Among the non-trivalent RE ions in GaN, the tetravalent Ce$^{4+}$, Pr$^{4+}$, and Tb$^{4+}$ are expected to be predominant over their trivalent counterparts in doped GaN samples prepared under or close to p-type conditions, i.e., when the Fermi level is closer to the VBM; see Fig.~\ref{fig;fe}(c). Ce$^{4+}$, in particular, has a very large range of the Fermi-level values, from $E_{\it v}$ to $E_{\it v} + 2.37$ eV, in which it is energetically more favorable than Ce$^{3+}$. The divalent Eu$^{2+}$ and Yb$^{2+}$, on the other hand, are expected to be predominant over the trivalent ions in samples prepared under n-type conditions. Note, however, that given the very small Yb$^{2+}$-favorable range that is very close to the CBM, see Fig.~\ref{fig;fe}(a), the divalent Yb$^{2+}$ is expected to be much harder to achieve during synthesis. It may, for example, be photogenerated under irradiation.

The stability of these non-trivalent RE ions was previously suggested by Dorenbos and van der Kolk on the basis of a semi-empirical model \cite{Dorenbos2006APL}. The authors fixed the Eu$^{2+}$ level at 3.1--3.2 eV above the VBM, which happens to coincide with the {\it thermodynamic} transition level $(0/-)$ of Eu$_{\rm Ga}$ we report earlier in Fig.~\ref{fig;fe}(b) and Table \ref{tab;re}. The Ce$^{3+}$, Pr$^{3+}$, Tb$^{3+}$, and Yb$^{2+}$ levels proposed in Ref.~\citenum{Dorenbos2006APL} are qualitatively consistent with our results for the $(+/0)$ level of Ce$_{\rm Ga}$, Pr$_{\rm Ga}$, and Tb$_{\rm Ga}$ and the $(0/-)$ level of Yb$_{\rm Ga}$, respectively; the difference is $\sim$0.5--0.8 eV. The semi-empirical model, however, offers no information on the $(+/0)$ level associated with the localized hole on the basal N atom we find in the other early and middle lanthanide defects. Through DFT-based calculations, Svane et al.~\cite{Svane2006} found (in self-interaction corrected, spin-polarized LDA calculations) that the $(0/-)$ level of $Ln_{\rm Ga}$ ($Ln$ = Nd, Pm, Sm, Eu, Ho, Er, Tm, and Yb) is above the band gap. Such a finding is in contrast to our results for $Ln$ = Eu and Yb, and not consistent with the fact that the $(0/-)$ level of the other $Ln_{\rm Ga}$ defects in the group is electronically unstable. Sanna et al.~\cite{Sanna2009} (who adopted an LDA$+$$U$ scheme within a density-functional-based tight-binding method), on the other hand, found the $(0/-)$ level of $Ln_{\rm Ga}$ ($Ln$ = Eu, Er, and Tm) to be within the host band gap, which is also in contrast to our results for $Ln$ = Er and Tm; their calculated level for $Ln$ = Eu is too far from the CBM and thus not consistent with experimental observations (see Ref.~\citenum{Hoang2021PRM} for a more detailed discussion). Note that our current results for $Ln$ = Eu and Er are in agreement with those we reported previously \cite{Hoang2021PRM,Hoang2015RRL}; small discrepancies, if present, can be due to the different versions of the PAW potentials used in the previous and current work. 

Experimentally, Eu is known to be mixed-valence in GaN, and significant Eu$^{2+}$ concentrations (e.g., $c({\rm Eu}^{2+})/c({\rm Eu}^{3+})>1$) have been achieved in GaN via co-doping with O and Si and tuning the growth conditions \cite{Mitchell2017MCP,Nunokawa2020JAP}. As we discussed in detail in Ref.~\citenum{Hoang2021PRM}, the O and Si co-doping, in which O$_{\rm N}$ and Si$_{\rm Ga}$ act as shallow donors, (i) shifts the Fermi level toward the CBM (``the global effect''), thus placing it in or close to the Fermi-level range in which Eu$^{2+}$ is energetically more favorable than Eu$^{3+}$, and (ii) extends the Eu$^{2+}$-favorable range via defect association (``the local effect''); the relatively low growth temperature also benefits high concentrations of defect complexes between Eu$_{\rm Ga}$ and O$_{\rm N}$ (or Si$_{\rm Ga}$) \cite{Hoang2021PRM}. We are not yet aware of any experimental report on the presence of Ce$^{4+}$, Pr$^{4+}$, Tb$^{4+}$, and Yb$^{2+}$ in GaN. 

In the following, we describe in detail the local lattice environment of the RE defects and analyze the electronic structure to understand why only certain defect configurations are electronically and energetically stable.

\subsection{Local lattice environment}\label{sec;latt}

\begin{figure}%[t]%
\vspace{0.2cm}
\includegraphics*[width=\linewidth]{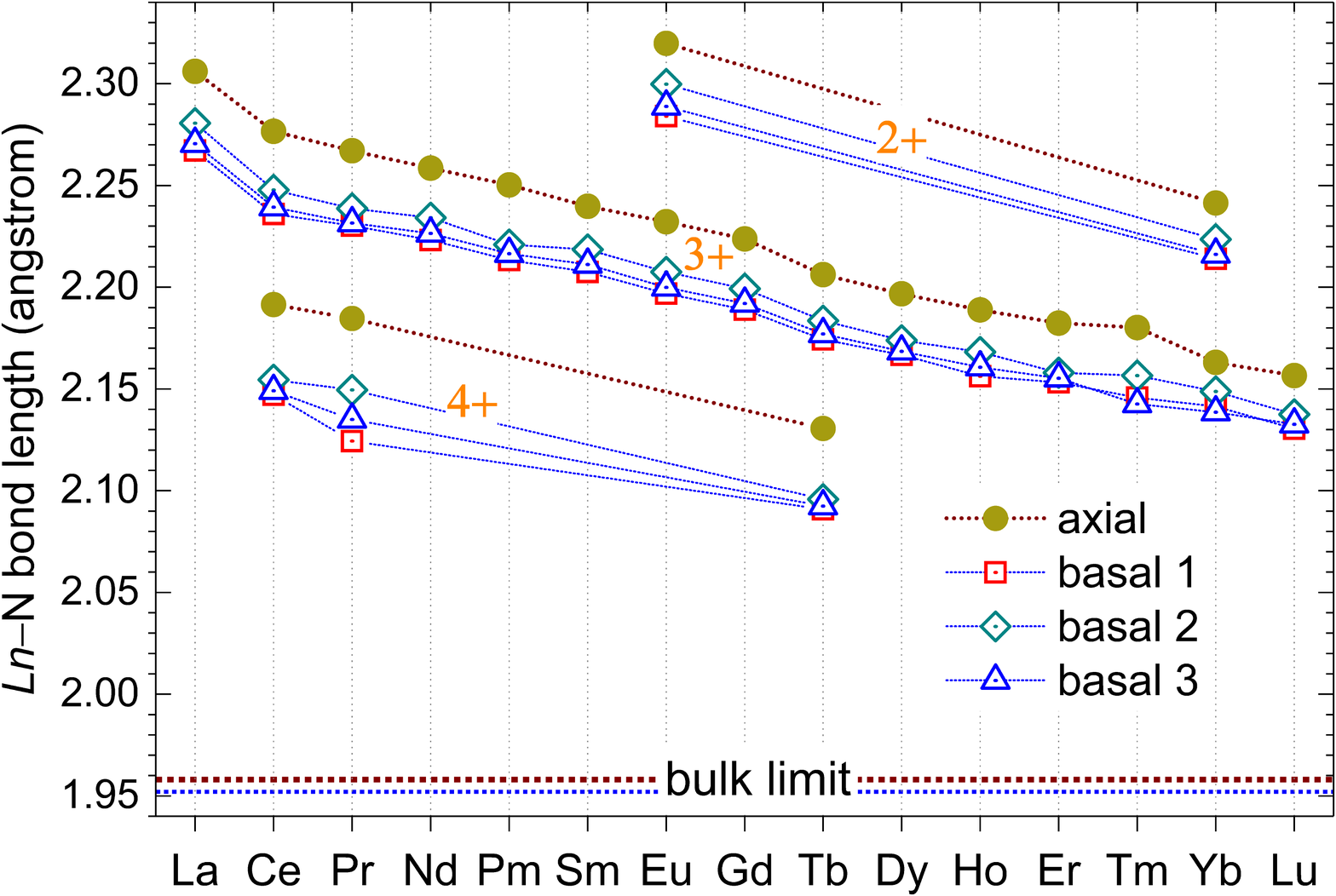}
\caption{Axial and basal $Ln$--N bond lengths (in {\AA}) in the isolated $Ln_{\rm Ga}^+$, $Ln_{\rm Ga}^0$, or $Ln_{\rm Ga}^-$ defect configuration. The valence of the RE ion ($2+$, $3+$, or $4+$) is indicated. The dotted lines connecting the symbols are to guide the eyes. The (dark red and blue) dotted lines near the bottom of the figure mark the axial and basal Ga--N bond length values in bulk GaN.}
\label{fig;bonds} 
\end{figure}

Figure \ref{fig;bonds} shows the $Ln$--N bond lengths in defect configurations $Ln_{\rm Ga}^+$, $Ln_{\rm Ga}^0$, and $Ln_{\rm Ga}^-$. For each charge state, we find that the calculated axial and basal bond lengths decrease monotonically as $Ln$ goes from La to Lu. Compared to the Ga--N bonds in bulk GaN, the $Ln$--N bonds are longer due to the outward relaxation of $Ln$'s neighboring N atoms. In addition, the difference between the axial and basal $Ln$--N bonds is larger and there is a small variation among the basal $Ln$--N bonds. The $Ln$ ion is slightly off-center. In the $Ln_{\rm Ga}^0$ configuration, for example, the $Ln^{3+}$ ion moves away from the original Ga site and predominantly toward the basal plane; the off-centering is smallest for $Ln$ = La ($\sim$0.03 {\AA}) and largest for  $Ln$ = Tm (0.08 {\AA}). The local distortion at the Ga site where the $Ln$ dopant is incorporated is thus more pronounced and slightly deviates from the $C_{3v}$ symmetry. Such significant local lattice distortion should relax the Laporte selection rules, making intra-$f$ optical transitions possible even for isolated RE centers in the host.

\begin{figure*}%[h]
\centering
\includegraphics[width=\linewidth]{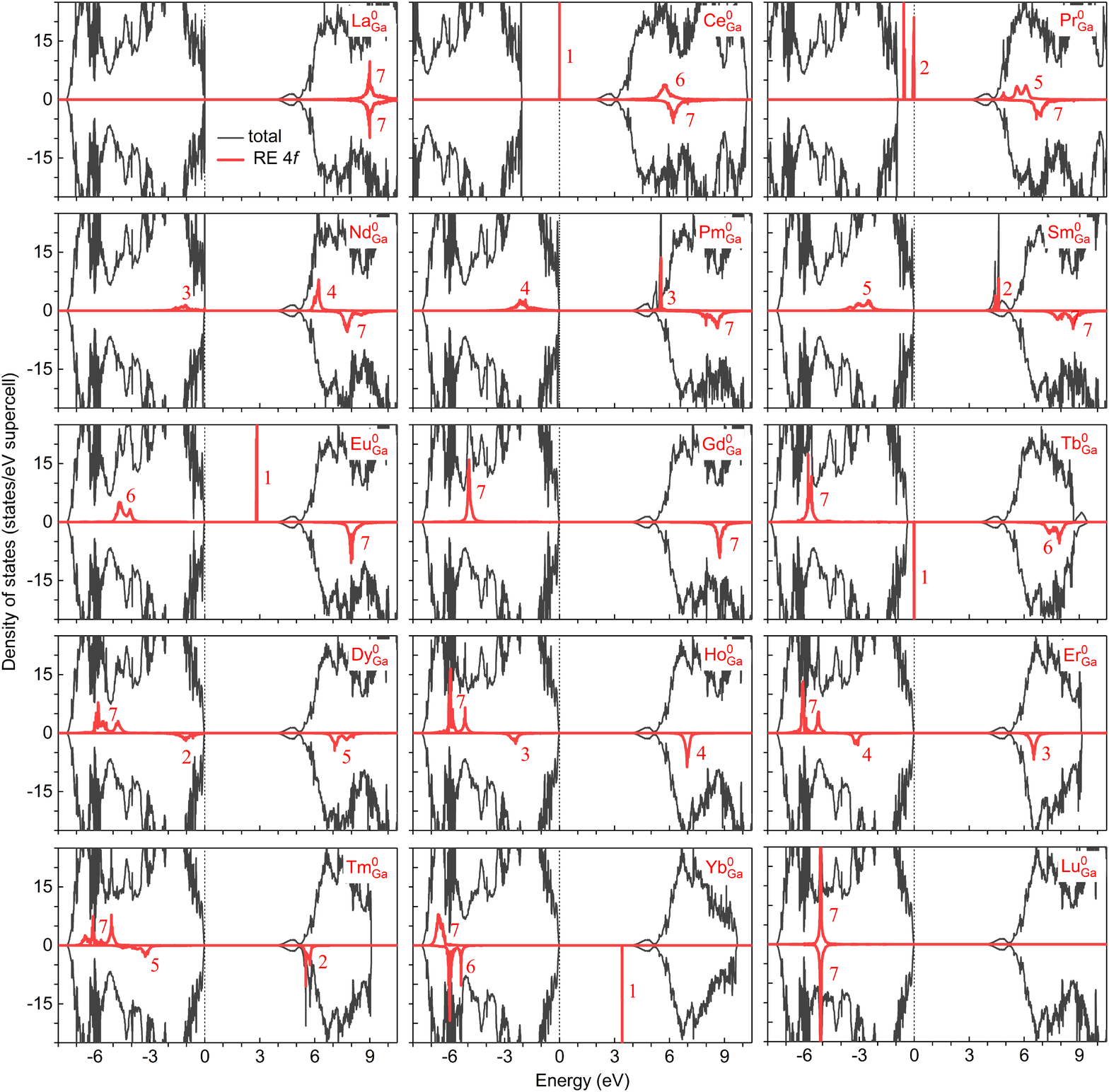}
\caption{Total and $Ln$ $4f$-projected densities of states (DOS) of $Ln$-doped GaN, i.e., the $Ln_{\rm Ga}^0$ defect configuration, obtained in HSE calculations. The spin-majority spectrum is on the $+y$ axis, and the spin-minority spectrum is on the $-y$ axis. The number of $Ln$ $4f$ electrons at the projected DOS peaks is indicated. The zero of energy is set to the highest occupied state.}
\label{fig;dos}
\end{figure*}

In the $Ln_{\rm Ga}^+$ configuration of group B defects (i.e., $Ln$ = La, Nd, Pm, Sm, Eu, Gd, and Dy), not included in Fig.~\ref{fig;bonds}, the local lattice distortion is a combination of those of the constituent defects ($Ln_{\rm Ga}^0$ and $h^\ast$). The presence of the localized hole on the N atom elongates the $Ln$--N bond, making that basal bond even longer than the axial $Ln$--N bond. The difference between the elongated basal $Ln$--N bond and the axial $Ln$--N bond is smallest for $Ln$ = Dy (0.01 {\AA}) and largest for $Ln$ = Eu (0.07 {\AA}). Due to the bond elongation, the other two basal $Ln$--N bonds are slightly shortened. The axial $Ln$--N bond length in these $Ln_{\rm Ga}^+$ defects are almost the same as that in $Ln_{\rm Ga}^0$.

\subsection{Electronic structure}\label{sec;elec}

Figure \ref{fig;dos} shows the electronic density of states (DOS) of $Ln$-doped GaN, specifically the $Ln_{\rm Ga}^0$ defect configuration described earlier. Note that by using the same 96-atom supercell model and thus a small ($\sim$2\%) dopant concentration, we avoid possible spurious $Ln$--$Ln$ interaction caused by the use of periodic boundary conditions and focus on the electronic structure of $Ln$-doped GaN at the dilute doping limit. We find that $Ln$ in $Ln_{\rm Ga}^0$ donates three outer electrons and becomes $Ln^{3+}$, consistent with our analysis in Sec.~\ref{sec;energetics}. The $Ln$ $4f$-projected DOS reveals the evolution of the electronic structure across the lanthanide series: Starting with La$_{\rm Ga}^0$ ($4f^0$) where the spin-up and spin-down $4f$ states are unoccupied and deep in the conduction band, these states move down toward the valence band one by one, first with the spin-up $4f$ states, as the number of the $4f$ electrons increases. In Gd$_{\rm Ga}^0$ ($4f^7$), there is a narrow peak with seven spin-up $4f$ electrons in the valence band and another with seven spin-down $4f$ electrons in the conduction band. From Tb$_{\rm Ga}^0$, the spin-down $4f$ states start moving toward the valence band until all the $4f$ states are occupied and deep in the valence band (in the case of Lu$_{\rm Ga}^0$, $4f^{14}$).

In addition to confirming the electronic stability of $Ln_{\rm Ga}^0$ (i.e., the trivalent $Ln^{3+}$) for all the elements in the lanthanide series, the calculated electronic structure also reveals if other charge states can be stabilized. We start with $Ln_{\rm Ga}^0$, $Ln$ = Ce, Pr, or Tb, which has the occupied $4f$ states in the host band gap. Upon removing one electron from this neutral configuration, the electron is removed from the highest occupied state (which is an $Ln$ $4f$ state). This results in $Ln^{3+}$ being oxidized to the tetravalent $Ln^{4+}$, thus explaining the stabilization of the $Ln_{\rm Ga}^+$ defects (i.e., $Ln^{4+}$ at the Ga site) in group C. Yb$_{\rm Ga}^0$ (group A) and Eu$_{\rm Ga}^0$ (group B) also have in-gap $4f$ states, but they are unoccupied. In this case, upon adding an electron to the neutral charge state, the electron is added to the lowest unoccupied state (an $Ln$ $4f$ state). This results in $Ln^{3+}$ being reduced to the divalent $Ln^{2+}$, thus explaining the stabilization of Yb$_{\rm Ga}^-$ and Eu$_{\rm Ga}^-$.

For all other $Ln_{\rm Ga}^0$ defect configurations whose electronic structure does not have $Ln$ $4f$ states in the host band gap, a true $Ln_{\rm Ga}^+$ or $Ln_{\rm Ga}^-$ charge state cannot be stabilized. This is because upon removing or adding an electron, the electron is removed from the VBM or added to the CBM that consists of delocalized host states. In other words, valence change cannot occur on the RE ion. Note that the nature of the removed electron (i.e., the electron hole) in the group B defects is different. Instead of being delocalized like in the late lanthanide (i.e., group A) defects, the hole is localized on a basal N atom in the case of $Ln_{\rm Ga}^+$ with $Ln$ = La, Nd, Pm, Sm, Eu, Gd, or Dy, due to strong $Ln$--N interaction. An examination of the electronic structure of the neutral charge state of these $Ln_{\rm Ga}^+$ defects shows that, indeed, there is stronger mixing between the N $2p$ states and the $Ln$ states at the VBM (not clearly seen in Fig.~\ref{fig;dos} due to the limited resolution) compared to that found in the late lanthanide defects. This explains why the configuration $Ln_{\rm Ga}^+ = Ln_{\rm Ga}^0 + h^\ast$ can be stabilized in group B, but not in group A. 

Overall, defect formation is determined by the electronic structure, as it has also been discussed in other classes of materials \cite{Hoang2018JPCM}. Through a careful examination and detailed discussion of the electronic structure of $Ln$-doped GaN, we explain why certain $Ln_{\rm Ga}$ defect configurations can be stabilized in GaN while others cannot. Further discussion of the electronic structure {\it vis-\`{a}-vis} defect formation in the case of $Ln$ = Eu and Er can be found in our previous work \cite{Hoang2016RRL,Hoang2021PRM}. It is important to note that the $Ln$-derived peaks in the DOS (Fig.~\ref{fig;dos}) are {\it not} defect energy levels associated with $Ln_{\rm Ga}$. Indeed, those Kohn-Sham levels cannot directly be identified with any defect levels that can be observed in experiments \cite{Freysoldt2014RMP}. The defect energy levels of $Ln_{\rm Ga}$ in the host band gap, if present, must be calculated using the total energies of the charge states of $Ln_{\rm Ga}$ as described and reported in Sec.~\ref{sec;energetics}.

Experimental data on the location of the RE $4f$ states in the electronic structure of RE-doped GaN has been scarce. Through resonant photoemission experiments on RE-doped GaN thin films, McHale et al.~\cite{McHale2011EPJAP} found that the occupied Gd, Er, and Yb $4f$ states are deep in the host's valence band, which is consistent with our results for Gd-, Er-, and Yb-doped GaN reported in Fig.~\ref{fig;dos}. 

\subsection{Defect-mediated optical transitions}\label{sec;opt}

\begin{figure}%[t]%
\vspace{0.2cm}
\includegraphics*[width=\linewidth]{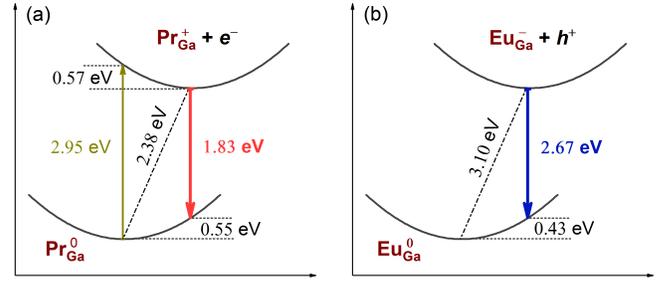}
\caption{Configuration-coordinate diagram illustrating optical absorption (up arrow) and emission (down arrow) processes involving (a) Pr$_{\rm Ga}$ and (b) Eu$_{\rm Ga}$ in GaN. The dash-dotted line indicates the thermal energy (i.e., ZPL). The values sandwiched between two dotted lines are the relaxation energies (i.e., the Franck-Condon shifts). Axes are not to scale.}
\label{fig;cc} 
\end{figure}

Like native defects and non-RE impurities that possess defect levels in the host band gap, certain isolated $Ln_{\rm Ga}$ defects in GaN can act as carrier traps in defect-to-band and band-to-defect transitions, including photoionization and radiative capture. Of these optical processes, those involving valence change on the $Ln$ ion have also been referred to as ``charge-transfer'' (CT) transitions in the literature (as opposed to the $4f$--$4f$ and $5d$--$4f$ transitions) \cite{Blasse1994Book}. Transitions involving the ($+/0$) level of group B defects, see Fig.~\ref{fig;fe}(b), are not strictly of the CT type as the trapped hole is localized at the N site and not on the $Ln$ ion. As discussed later, the $Ln_{\rm Ga}$ defects can also act as carrier traps for the intra-$f$ luminescence. 

\begin{table}%[b]
\caption{Optical transitions associated with $Ln_{\rm Ga}$ defects in GaN. The right (left) arrows are for the absorption (emission) processes; Y(es) and N(o) are used to indicate whether or not the transitions are of the ``charge-transfer'' (CT) type. The thermal ($E_{\rm therm}$), absorption ($E_{\rm abs}$), and emission ($E_{\rm em}$) energies are all in eV. $S_{\rm \{e,g\}}$ are the estimated Huang-Rhys factors; see the text. Absorption peaks that fall outside the host band gap are also included (and italicized).}\label{tab;opt}
\begin{center}
\begin{ruledtabular}
\begin{tabular}{lcccccc}
Optical transition & CT & $E_{\rm therm}$ & $E_{\rm abs}$ & $S_{\rm e}$ & $E_{\rm em}$ & $S_{\rm g}$ \\
\colrule
La$_{\rm Ga}^0$ $\rightleftharpoons$ La$_{\rm Ga}^+$ $+$ $e^-$ & N & 3.06 & {\it 3.54} & 16.1 & 2.53 & 17.7 \\
Ce$_{\rm Ga}^0$ $\rightleftharpoons$ Ce$_{\rm Ga}^+$ $+$ $e^-$ & Y & 1.16 & 1.70 & 18.2 & 0.69 & 15.6 \\
Ce$_{\rm Ga}^+$ $\rightleftharpoons$ Ce$_{\rm Ga}^0$ $+$ $h^+$ & Y & 2.37 & 2.84 & 15.6 & 1.83 & 18.2 \\
Pr$_{\rm Ga}^0$ $\rightleftharpoons$ Pr$_{\rm Ga}^+$ $+$ $e^-$ & Y & 2.38 & 2.95 & 19.0 & 1.83 & 18.2 \\
Pr$_{\rm Ga}^+$ $\rightleftharpoons$ Pr$_{\rm Ga}^0$ $+$ $h^+$ & Y & 1.16 & 1.70 & 18.2 & 0.59 & 19.0 \\
Nd$_{\rm Ga}^0$ $\rightleftharpoons$ Nd$_{\rm Ga}^+$ $+$ $e^-$ & N & 3.23 & {\it 3.59} & 11.9 & 2.70 & 17.7 \\
Pm$_{\rm Ga}^0$ $\rightleftharpoons$ Pm$_{\rm Ga}^+$ $+$ $e^-$ & N & 3.25 & {\it 3.74} & 16.3 & 2.73 & 17.2 \\
Sm$_{\rm Ga}^0$ $\rightleftharpoons$ Sm$_{\rm Ga}^+$ $+$ $e^-$ & N & 3.37 & {\it 3.74} & 12.2 & 2.88 & 16.1 \\
Eu$_{\rm Ga}^0$ $\rightleftharpoons$ Eu$_{\rm Ga}^+$ $+$ $e^-$ & N & 3.32 & {\it 3.75} & 14.2 & 2.83 & 16.3 \\
Eu$_{\rm Ga}^0$ $\rightleftharpoons$ Eu$_{\rm Ga}^-$ $+$ $h^+$ & Y & 3.10 & {\it 3.93} & 27.6 & 2.67 & 14.5 \\
Gd$_{\rm Ga}^0$ $\rightleftharpoons$ Gd$_{\rm Ga}^+$ $+$ $e^-$ & N & 3.45 & {\it 3.75} & 10.0 & 2.94 & 17.1 \\
Tb$_{\rm Ga}^0$ $\rightleftharpoons$ Tb$_{\rm Ga}^+$ $+$ $e^-$ & Y & 2.99 & 3.49 & 16.7 & 2.54 & 14.9 \\
Dy$_{\rm Ga}^0$ $\rightleftharpoons$ Dy$_{\rm Ga}^+$ $+$ $e^-$ & N & 3.40 & {\it 3.75} & 12.0 & 2.94 & 15.3 \\
Yb$_{\rm Ga}^0$ $\rightleftharpoons$ Yb$_{\rm Ga}^-$ $+$ $h^+$ & Y & 3.48 & {\it 3.87} & 13.0 & 3.15 & 10.8 \\
\end{tabular}
\end{ruledtabular}
\end{center}
\begin{flushleft}
\end{flushleft}
\end{table} 

Figure \ref{fig;cc} shows examples of absorption and emission transitions involving Pr$_{\rm Ga}$ and Eu$_{\rm Ga}$. Under illumination, for example, the isolated Pr$_{\rm Ga}^0$ can absorb a photon and become ionized to Pr$_{\rm Ga}^+$ with the removed electron being excited into the conduction band. The {\it peak} absorption energy ($E_{\rm abs}$) corresponding to the optical transition level $E_{\rm opt}^{0/+}$ (i.e., the formation energy difference between Pr$_{\rm Ga}^0$ and Pr$_{\rm Ga}^+$ in the lattice configuration of Pr$_{\rm Ga}^0$) is calculated to be 2.95 eV, with a relaxation energy (i.e., the Franck-Condon shift in the excited state, $d_{\rm FC}^{\rm e}$) of 0.57 eV. Pr$_{\rm Ga}^+$ can then capture an electron from the CBM (e.g., previously excited from Pr$_{\rm Ga}^0$ to the conduction band) or from a shallow donor level and emit a photon; here, we assume that the recombination is radiative. The {\it peak} emission energy ($E_{\rm em}$) corresponding to the optical transition level $E_{\rm opt}^{+/0}$ (i.e., the formation energy difference between Pr$_{\rm Ga}^+$ and Pr$_{\rm Ga}^0$ in the lattice configuration of Pr$_{\rm Ga}^+$) is 1.83 eV, with a relaxation energy (i.e., the Franck-Condon shift in the ground state, $d_{\rm FC}^{\rm g}$) of 0.55 eV; see Fig.~\ref{fig;cc}(a). The thermal energy [$E_{\rm therm}$, also referred to as the zero-phonon line (ZPL) energy] of the Pr$_{\rm Ga}^0$ $\rightleftharpoons$ Pr$_{\rm Ga}^+$ $+$ $e^-$ transitions is 2.38 eV, related to the thermodynamic transition level $\epsilon(+/0)$ of Pr$_{\rm Ga}$. The ZPL marks the initial {\it onset} of the absorption band. Transitions between the $(+/0)$ level of Pr$_{\rm Ga}$ and an electron hole at the VBM, i.e., Pr$_{\rm Ga}^+$ $\rightleftharpoons$ Pr$_{\rm Ga}^0$ $+$ $h^+$, are also possible and would lead to a different set of the thermal, absorption, and emission energies as seen in Table \ref{tab;opt}. In the case of Eu$_{\rm Ga}$, the emission process involves Eu$_{\rm Ga}^-$ capturing an electron hole either from the VBM or some shallow acceptor level; assuming that the recombination is radiative, the peak emission energy is calculated to be 2.67 eV, with a relaxation energy ($d_{\rm FC}^{\rm g}$) of 0.43 eV; see Fig.~\ref{fig;cc}(b). The thermal energy of the Eu$_{\rm Ga}^0$ $\rightleftharpoons$ Eu$_{\rm Ga}^-$ $+$ $h^+$ transitions is 3.10 eV, related to the thermodynamic transition level $\epsilon(0/-)$ of Eu$_{\rm Ga}$. Transitions involving the $(+/0)$ level of Eu$_{\rm Ga}$ are also possible; see Table \ref{tab;opt}. 

Optical transitions involving the other RE defects with in-gap levels are investigated similarly, and all the results are listed in Table \ref{tab;opt}. The Franck-Condon shifts ($d_{\rm FC}^{\rm e}$ and $d_{\rm FC}^{\rm g}$) can be obtained from the reported values for the thermal ($E_{\rm therm}$), absorption ($E_{\rm abs}$), and emission ($E_{\rm em}$) energies using the following relations \cite{Alkauskas2016JAP}
\begin{align}
	E_{\rm abs} = E_{\rm therm} + d_{\rm FC}^{\rm e},\\
	E_{\rm em} = E_{\rm therm} - d_{\rm FC}^{\rm g}.
\end{align}
The Stokes shift, i.e., the difference between the absorption and emission energies, is the sum of the Franck-Condon shifts in the excited and ground states \cite{Alkauskas2016JAP}:
\begin{equation}
 E_{\rm abs} - E_{\rm em} = d_{\rm FC}^{\rm e} + d_{\rm FC}^{\rm g}.
\end{equation}
Note that for optical processes involving exchange of electrons (holes) with the CBM (VBM), the thermal, absorption, and emission energies are measured relative to the CBM (VBM). For all the processes listed in Table \ref{tab;opt}, we find that $d_{\rm FC}^{\rm \{e,g\}} =$ 0.30--0.83 eV. Given the rather large calculated relaxation energies, the absorption and emission are expected to be broad. For comparison, Dorenbos \cite{Dorenbos2017OM} estimated (semi-empirically) that the relaxation energy is of the order of 0.6 eV for CT transitions in various RE-doped materials. And for transition-metal defects in GaN, Wickramaratne et al.~\cite{Wickramaratne2019PRB} reported $d_{\rm FC}^{\rm \{e,g\}} =$ 0.32--0.40 eV for optical processes involving Fe$_{\rm Ga}$ defects. 

The Huang-Rhys (HR) factor \cite{Huang1950}, which characterizes the electron-phonon coupling strength, is given by \cite{Alkauskas2016JAP}
\begin{equation}
 S_{\rm \{e,g\}} = \frac{d_{\rm FC}^{\rm \{e,g\}}}{\hbar \omega_{\rm \{e,g\}}},
\end{equation}     
where $\omega_{\rm e}$ and $\omega_{\rm g}$ are the effective phonon frequencies in the excited and ground state. If we assume $\hbar \omega_{\rm e} = \hbar \omega_{\rm g} = 30$ meV (a typical phonon frequency in GaN \cite{Alkauskas2016JAP}), the HR factors are {\it estimated} to be $S_{\rm \{e,g\}} =$ $10.0$--$27.6$; see Table \ref{tab;opt}. With such large HR factors ($S_{\rm \{e,g\}} \gg 1$), the defects can be considered as having large electron-phonon coupling. In this case, the peak absorption or emission energy coincides with the optical transition level $E_{\rm opt}^{q/q'}$ \cite{Alkauskas2012PRL,Alkauskas2016JAP}, thus justifying our earlier peak assignment. 

\begin{figure}%[t]%
\vspace{0.2cm}
\includegraphics*[width=\linewidth]{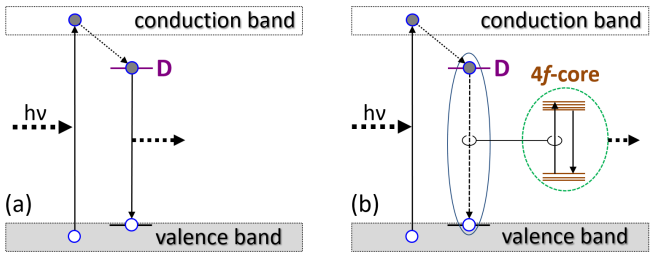}
\caption{Schematic illustration of possible optical processes involving a $Ln_{\rm Ga}$ defect with an in-gap energy level (D) in GaN following a band-to-band excitation of the host. The recombination of the excited electron trapped at D and a free hole can be (a) radiative or (b) nonradiative; see the text. Optical processes involving hole trapping are similar.}
\label{fig;opt} 
\end{figure}

It should be noted that for all the emission processes listed in Table \ref{tab;opt} we assume {\it radiative} recombination of the trapped electron (hole) and the free hole (electron). This is illustrated in Fig.~\ref{fig;opt}(a) for the case of electron trapping. For CT-type processes, however, the CT emission (which is the reverse of the CT absorption) may not be observed. This is because the recombination energy can quickly be absorbed by the $4f$-electron core of the $Ln$ ion. Figure~\ref{fig;opt}(b) illustrates such a process where the trapped electron recombines {\it nonradiatively} with a free hole, and the recombination energy is transferred to the $4f$-electron core which then excites the $Ln$ ion and leads to intra-$f$ luminescence (not explicitly considered in this work). The competition between the two mechanisms illustrated in Fig.~\ref{fig;opt} is expected to be dependent on specific defect configurations, including the energy difference between the carrier trap level (D) and the excited $4f$ states \cite{vanPieterson2000JL}. That should apply to defects not just of CT-type transitions but also of non-CT type, including both isolated $Ln$ defects and $Ln$-related defect complexes (such as those Eu-related complexes reported in Ref.~\citenum{Hoang2021PRM}).   

Let us take the Eu$_{\rm Ga}^-$ $+$ $h^+$ $\rightarrow$ Eu$_{\rm Ga}^0$ transition as an example. After a nonradiative recombination of the electron (trapped at Eu$_{\rm Ga}^-$) and a free hole ($h^+$), the defect becomes Eu$_{\rm Ga}^0$ with an electron being promoted from the ground $^7F_J$ state to the excited $^5D_J$ state of the Eu$^{3+}$ $4f$ manifold. The subsequent relaxation from the excited $^5D_J$ state to the ground state would result in a sharp red luminescence, as opposed to a broad blue luminescence as one would observe with the CT emission illustrated in Fig.~\ref{fig;cc}(b). For further discussion of the role of Eu$_{\rm Ga}$ and Eu-related defect complexes as carrier traps for Eu$^{3+}$ intra-$f$ luminescence in GaN, see Ref.~\citenum{Hoang2021PRM}.

Experimentally, although defect-to-band and band-to-defect optical transitions in RE-doped GaN likely affect the performance of the material, they have apparently not been well discussed, except probably in the case of $Ln = $Eu. There have been reports of a Eu-related, broad CT excitation band centered at about 3.0--3.2 eV above the VBM in the photoluminescence excitation (PLE) spectra of Eu-doped GaN \cite{Morishima1999PSS,Tanaka2003PSS,Nyein2003APL,Higuchi2010PRB} or of a CT absorption peak at 0.37 eV below the CBM \cite{Li2002JCG,Sawahata2005STAM}. The excitation band appears to largely overlap with the host lattice excitation band \cite{Morishima1999PSS,Tanaka2003PSS,Sawahata2005STAM}. The initial onset of the excitation band is at about 2.9--3.0 eV \cite{Morishima1999PSS,Tanaka2003PSS,Nyein2003APL,Sawahata2005STAM}, which is in reasonable agreement with the ZPL ($E_{\rm therm} = 3.10$ eV) obtained for the Eu$_{\rm Ga}^0$ $\rightarrow$ Eu$_{\rm Ga}^-$ $+$ $h^+$ transition (see Table \ref{tab;opt}) and consistent with the presence of the defect level ($0/-$) of Eu$_{\rm Ga}$ discussed in Sec.~\ref{sec;energetics}. 

There have been no experimental reports of CT emission in Eu-doped GaN. This may suggest that the mechanism illustrated in Fig.~\ref{fig;cc}(b) is predominant. Indeed, the mentioned defect level has been thought to play an important role in the Eu$^{3+}$ intra-$f$ luminescence \cite{Li2002JCG,Tanaka2003PSS,Nyein2003APL,Sawahata2005STAM,Higuchi2010PRB}. The isolated Eu$_{\rm Ga}$ defect is unlikely the only luminescent Eu$^{3+}$ center, however, and there may be Eu-related defect complexes in Eu-doped GaN samples that are more efficient for nonresonant excitation of Eu$^{3+}$ \cite{Hoang2021PRM}.  

Finally, it should be noted that although CT emission appears to be rare \cite{vanPieterson2000JL,Dorenbos2017OM}, it has been observed in various Yb-doped materials \cite{Nakazawa1978CPL,Nakazawa1979JL,vanPieterson2000JL} and in Sr$_2$CeO$_4$ \cite{Danielson1998Science}.  

\section{Conclusions and outlook}

We have carried out a systematic study of lanthanide (La--Lu) defects in GaN using hybrid density-functional defect calculations. We find that all the $Ln$ dopants, when incorporated into the host material at the Ga site ($Ln_{\rm Ga}$), are stable as trivalent ions. In addition to the trivalent state, Eu and Yb are also electronically stable as divalent and Ce, Pr, and Tb as tetravalent. The mixed-valence dopants are characterized by having unoccupied (Eu and Yb) or occupied (Ce, Pr, and Tb) $4f$ states in the host band gap and possessing defect levels that are associated with valence change on the $Ln$ ion. The early and middle lanthanide (La--Dy) dopants, except those (Ce, Pr, and Tb) that can be stabilized in the tetravalent state, introduce a defect level just above the VBM. This level is not associated with valence change on the $Ln$ ion but with the formation of a localized hole on the N atom basally bonded to $Ln$. That localized state (and hence the defect level) is absent in the late lanthanide (Ho--Lu) defects due to the weaker $Ln$--N interaction. The location of the $Ln$-related defect energy levels and the $Ln$ $4f$ states in the energy spectrum of the host material thus has been now determined from first principles. We also find that all the $Ln$ defects significantly distort the local lattice environment, thus relaxing the selection rules and allowing for parity-forbidden intra-$f$ transitions.

The optical properties are investigated by considering band-to-defect and defect-to-band optical transitions involving the RE defects. We find that the isolated $Ln_{\rm Ga}$ defects (except $Ln =$ Ho, Er, Tm, and Lu with no localized in-gap levels) can be the source of broad absorption and emission bands. The emission bands, especially those of ``charge-transfer'' type, however, may not be observed due to a competing mechanism in which the recombination energy is transferred into the $4f$-electron core of the $Ln$ ion. The defects thus can also act as carrier traps for intra-$f$ luminescence through nonresonant excitation of the $Ln$ ion. Further computational and experimental studies are needed to characterize these transitions and to better understand their impact on the performance of the material. These may include first-principles calculations of photoionization and carrier capture rates \cite{Razinkovas2021PRB,Dreyer2020PRB} which can provide a more quantitative understanding.

Finally, the results reported in this work can serve as the benchmark for calculations using computationally light--and often with limited predictive power--methods such as DFT$+$$U$ as well as more compute-intensive, post-DFT approaches. They also form the basis for further studies of RE-related defects in GaN, including direct interaction between the $Ln$ dopant and native defects and/or impurities that may be present in the host material. As seen in the case of Er \cite{Hoang2015RRL,Hoang2016RRL} and Eu \cite{Hoang2021PRM} dopants in GaN, defect association can significantly modify the electronic behavior of a defect and may thus offer interesting physics useful for electrical and optical control.

\begin{acknowledgments}

This work made use of resources in the Center for Computationally Assisted Science and Technology (CCAST) at North Dakota State University, which were made possible in part by NSF MRI Award No.~2019077.

\end{acknowledgments}

%\bibliography{../optoelectronics_refs}

\begin{thebibliography}{51}%
\makeatletter
\providecommand \@ifxundefined [1]{%
 \@ifx{#1\undefined}
}%
\providecommand \@ifnum [1]{%
 \ifnum #1\expandafter \@firstoftwo
 \else \expandafter \@secondoftwo
 \fi
}%
\providecommand \@ifx [1]{%
 \ifx #1\expandafter \@firstoftwo
 \else \expandafter \@secondoftwo
 \fi
}%
\providecommand \natexlab [1]{#1}%
\providecommand \enquote  [1]{``#1''}%
\providecommand \bibnamefont  [1]{#1}%
\providecommand \bibfnamefont [1]{#1}%
\providecommand \citenamefont [1]{#1}%
\providecommand \href@noop [0]{\@secondoftwo}%
\providecommand \href [0]{\begingroup \@sanitize@url \@href}%
\providecommand \@href[1]{\@@startlink{#1}\@@href}%
\providecommand \@@href[1]{\endgroup#1\@@endlink}%
\providecommand \@sanitize@url [0]{\catcode `\\12\catcode `\$12\catcode
  `\&12\catcode `\#12\catcode `\^12\catcode `\_12\catcode `\%12\relax}%
\providecommand \@@startlink[1]{}%
\providecommand \@@endlink[0]{}%
\providecommand \url  [0]{\begingroup\@sanitize@url \@url }%
\providecommand \@url [1]{\endgroup\@href {#1}{\urlprefix }}%
\providecommand \urlprefix  [0]{URL }%
\providecommand \Eprint [0]{\href }%
\providecommand \doibase [0]{https://doi.org/}%
\providecommand \selectlanguage [0]{\@gobble}%
\providecommand \bibinfo  [0]{\@secondoftwo}%
\providecommand \bibfield  [0]{\@secondoftwo}%
\providecommand \translation [1]{[#1]}%
\providecommand \BibitemOpen [0]{}%
\providecommand \bibitemStop [0]{}%
\providecommand \bibitemNoStop [0]{.\EOS\space}%
\providecommand \EOS [0]{\spacefactor3000\relax}%
\providecommand \BibitemShut  [1]{\csname bibitem#1\endcsname}%
\let\auto@bib@innerbib\@empty
%</preamble>
\bibitem [{\citenamefont {O'Donnell}\ and\ \citenamefont
  {Dierolf}(2010)}]{ODonnell2010Book}%
  \BibitemOpen
  \bibinfo {editor} {\bibfnamefont {K.}~\bibnamefont {O'Donnell}}\ and\
  \bibinfo {editor} {\bibfnamefont {V.}~\bibnamefont {Dierolf}},\ eds.,\
  \href@noop {} {\emph {\bibinfo {title} {{Rare Earth Doped III-Nitrides for
  Optoelectronic and Spintronic Applications}}}},\ \bibinfo {series} {Topics in
  Applied Physics}, Vol.\ \bibinfo {volume} {124}\ (\bibinfo  {publisher}
  {Springer},\ \bibinfo {address} {Dordrecht},\ \bibinfo {year}
  {2010})\BibitemShut {NoStop}%
\bibitem [{\citenamefont {Steckl}\ and\ \citenamefont
  {Zavada}(1999)}]{StecklMRS1999}%
  \BibitemOpen
  \bibfield  {author} {\bibinfo {author} {\bibfnamefont {A.}~\bibnamefont
  {Steckl}}\ and\ \bibinfo {author} {\bibfnamefont {J.}~\bibnamefont
  {Zavada}},\ }\bibfield  {title} {\bibinfo {title} {{Optoelectronic Properties
  and Applications of Rare-Earth-Doped GaN}},\ }\href
  {https://doi.org/10.1557/S0883769400053045} {\bibfield  {journal} {\bibinfo
  {journal} {MRS Bull.}\ }\textbf {\bibinfo {volume} {24}},\ \bibinfo {pages}
  {33} (\bibinfo {year} {1999})}\BibitemShut {NoStop}%
\bibitem [{\citenamefont {Blasse}\ and\ \citenamefont
  {Grabmaier}(1994)}]{Blasse1994Book}%
  \BibitemOpen
  \bibfield  {author} {\bibinfo {author} {\bibfnamefont {G.}~\bibnamefont
  {Blasse}}\ and\ \bibinfo {author} {\bibfnamefont {B.~C.}\ \bibnamefont
  {Grabmaier}},\ }\href@noop {} {\emph {\bibinfo {title} {{Luminescent
  Materials}}}}\ (\bibinfo  {publisher} {Springer-Verlag},\ \bibinfo {address}
  {Berlin},\ \bibinfo {year} {1994})\BibitemShut {NoStop}%
\bibitem [{\citenamefont {Weber}\ \emph {et~al.}(2010)\citenamefont {Weber},
  \citenamefont {Koehl}, \citenamefont {Varley}, \citenamefont {Janotti},
  \citenamefont {Buckley}, \citenamefont {Van~de Walle},\ and\ \citenamefont
  {Awschalom}}]{Weber2010PNAS}%
  \BibitemOpen
  \bibfield  {author} {\bibinfo {author} {\bibfnamefont {J.~R.}\ \bibnamefont
  {Weber}}, \bibinfo {author} {\bibfnamefont {W.~F.}\ \bibnamefont {Koehl}},
  \bibinfo {author} {\bibfnamefont {J.~B.}\ \bibnamefont {Varley}}, \bibinfo
  {author} {\bibfnamefont {A.}~\bibnamefont {Janotti}}, \bibinfo {author}
  {\bibfnamefont {B.~B.}\ \bibnamefont {Buckley}}, \bibinfo {author}
  {\bibfnamefont {C.~G.}\ \bibnamefont {Van~de Walle}},\ and\ \bibinfo {author}
  {\bibfnamefont {D.~D.}\ \bibnamefont {Awschalom}},\ }\bibfield  {title}
  {\bibinfo {title} {Quantum computing with defects},\ }\href
  {https://doi.org/10.1073/pnas.1003052107} {\bibfield  {journal} {\bibinfo
  {journal} {Proc. Natl. Acad. Sci.}\ }\textbf {\bibinfo {volume} {107}},\
  \bibinfo {pages} {8513} (\bibinfo {year} {2010})}\BibitemShut {NoStop}%
\bibitem [{\citenamefont {Gordon}\ \emph {et~al.}(2013)\citenamefont {Gordon},
  \citenamefont {Weber}, \citenamefont {Varley}, \citenamefont {Janotti},
  \citenamefont {Awschalom},\ and\ \citenamefont {Van~de
  Walle}}]{Gordon2013MRSBull}%
  \BibitemOpen
  \bibfield  {author} {\bibinfo {author} {\bibfnamefont {L.}~\bibnamefont
  {Gordon}}, \bibinfo {author} {\bibfnamefont {J.~R.}\ \bibnamefont {Weber}},
  \bibinfo {author} {\bibfnamefont {J.~B.}\ \bibnamefont {Varley}}, \bibinfo
  {author} {\bibfnamefont {A.}~\bibnamefont {Janotti}}, \bibinfo {author}
  {\bibfnamefont {D.~D.}\ \bibnamefont {Awschalom}},\ and\ \bibinfo {author}
  {\bibfnamefont {C.~G.}\ \bibnamefont {Van~de Walle}},\ }\bibfield  {title}
  {\bibinfo {title} {Quantum computing with defects},\ }\href
  {https://doi.org/10.1557/mrs.2013.206} {\bibfield  {journal} {\bibinfo
  {journal} {MRS Bull.}\ }\textbf {\bibinfo {volume} {38}},\ \bibinfo {pages}
  {802-807} (\bibinfo {year} {2013})}\BibitemShut {NoStop}%
\bibitem [{\citenamefont {Thiel}\ \emph {et~al.}(2011)\citenamefont {Thiel},
  \citenamefont {B\"{o}ttger},\ and\ \citenamefont {Cone}}]{Thiel2011JL}%
  \BibitemOpen
  \bibfield  {author} {\bibinfo {author} {\bibfnamefont {C.}~\bibnamefont
  {Thiel}}, \bibinfo {author} {\bibfnamefont {T.}~\bibnamefont {Böttger}},\
  and\ \bibinfo {author} {\bibfnamefont {R.}~\bibnamefont {Cone}},\ }\bibfield
  {title} {\bibinfo {title} {Rare-earth-doped materials for applications in
  quantum information storage and signal processing},\ }\href
  {https://doi.org/10.1016/j.jlumin.2010.12.015} {\bibfield  {journal}
  {\bibinfo  {journal} {J. Lumin.}\ }\textbf {\bibinfo {volume} {131}},\
  \bibinfo {pages} {353} (\bibinfo {year} {2011})}\BibitemShut {NoStop}%
\bibitem [{\citenamefont {Kunkel}\ and\ \citenamefont
  {Goldner}(2018)}]{Kunkel2018ZAAC}%
  \BibitemOpen
  \bibfield  {author} {\bibinfo {author} {\bibfnamefont {N.}~\bibnamefont
  {Kunkel}}\ and\ \bibinfo {author} {\bibfnamefont {P.}~\bibnamefont
  {Goldner}},\ }\bibfield  {title} {\bibinfo {title} {{Recent Advances in Rare
  Earth Doped Inorganic Crystalline Materials for Quantum Information
  Processing}},\ }\href {https://doi.org/10.1002/zaac.201700425} {\bibfield
  {journal} {\bibinfo  {journal} {Z. Anorg. Allg. Chem.}\ }\textbf {\bibinfo
  {volume} {644}},\ \bibinfo {pages} {66} (\bibinfo {year} {2018})}\BibitemShut
  {NoStop}%
\bibitem [{\citenamefont {Zhong}\ and\ \citenamefont
  {Goldner}(2019)}]{Zhong2019NP}%
  \BibitemOpen
  \bibfield  {author} {\bibinfo {author} {\bibfnamefont {T.}~\bibnamefont
  {Zhong}}\ and\ \bibinfo {author} {\bibfnamefont {P.}~\bibnamefont
  {Goldner}},\ }\bibfield  {title} {\bibinfo {title} {Emerging rare-earth doped
  material platforms for quantum nanophotonics},\ }\href
  {https://doi.org/10.1515/nanoph-2019-0185} {\bibfield  {journal} {\bibinfo
  {journal} {Nanophotonics}\ }\textbf {\bibinfo {volume} {8}},\ \bibinfo
  {pages} {2003} (\bibinfo {year} {2019})}\BibitemShut {NoStop}%
\bibitem [{\citenamefont {Mitchell}\ \emph {et~al.}(2021)\citenamefont
  {Mitchell}, \citenamefont {Austin}, \citenamefont {Timmerman}, \citenamefont
  {Dierolf},\ and\ \citenamefont {Fujiwara}}]{Mitchell2021NP}%
  \BibitemOpen
  \bibfield  {author} {\bibinfo {author} {\bibfnamefont {B.}~\bibnamefont
  {Mitchell}}, \bibinfo {author} {\bibfnamefont {H.}~\bibnamefont {Austin}},
  \bibinfo {author} {\bibfnamefont {D.}~\bibnamefont {Timmerman}}, \bibinfo
  {author} {\bibfnamefont {V.}~\bibnamefont {Dierolf}},\ and\ \bibinfo {author}
  {\bibfnamefont {Y.}~\bibnamefont {Fujiwara}},\ }\bibfield  {title} {\bibinfo
  {title} {Temporally modulated energy shuffling in highly interconnected
  nanosystems},\ }\href {https://doi.org/10.1515/nanoph-2020-0484} {\bibfield
  {journal} {\bibinfo  {journal} {Nanophotonics}\ }\textbf {\bibinfo {volume}
  {10}},\ \bibinfo {pages} {851} (\bibinfo {year} {2021})}\BibitemShut
  {NoStop}%
\bibitem [{\citenamefont {McHale}\ \emph {et~al.}(2011)\citenamefont {McHale},
  \citenamefont {McClory}, \citenamefont {Petrosky}, \citenamefont {Wu},
  \citenamefont {Palai}, \citenamefont {Losovyj},\ and\ \citenamefont
  {Dowben}}]{McHale2011EPJAP}%
  \BibitemOpen
  \bibfield  {author} {\bibinfo {author} {\bibfnamefont {S.}~\bibnamefont
  {McHale}}, \bibinfo {author} {\bibfnamefont {J.}~\bibnamefont {McClory}},
  \bibinfo {author} {\bibfnamefont {J.}~\bibnamefont {Petrosky}}, \bibinfo
  {author} {\bibfnamefont {J.}~\bibnamefont {Wu}}, \bibinfo {author}
  {\bibfnamefont {R.}~\bibnamefont {Palai}}, \bibinfo {author} {\bibfnamefont
  {Y.}~\bibnamefont {Losovyj}},\ and\ \bibinfo {author} {\bibfnamefont
  {P.}~\bibnamefont {Dowben}},\ }\bibfield  {title} {\bibinfo {title} {Resonant
  photoemission of rare earth doped {GaN} thin films},\ }\href
  {https://doi.org/10.1051/epjap/2011110235} {\bibfield  {journal} {\bibinfo
  {journal} {Eur. Phys. J. Appl. Phys.}\ }\textbf {\bibinfo {volume} {56}},\
  \bibinfo {pages} {11301} (\bibinfo {year} {2011})}\BibitemShut {NoStop}%
\bibitem [{\citenamefont {Morishima}\ \emph {et~al.}(1999)\citenamefont
  {Morishima}, \citenamefont {Maruyama}, \citenamefont {Tanaka},\ and\
  \citenamefont {Akimoto}}]{Morishima1999PSS}%
  \BibitemOpen
  \bibfield  {author} {\bibinfo {author} {\bibfnamefont {S.}~\bibnamefont
  {Morishima}}, \bibinfo {author} {\bibfnamefont {T.}~\bibnamefont {Maruyama}},
  \bibinfo {author} {\bibfnamefont {M.}~\bibnamefont {Tanaka}},\ and\ \bibinfo
  {author} {\bibfnamefont {K.}~\bibnamefont {Akimoto}},\ }\bibfield  {title}
  {\bibinfo {title} {{Growth of Eu Doped GaN and Electroluminescence from MIS
  Structure}},\ }\href@noop {} {\bibfield  {journal} {\bibinfo  {journal}
  {phys. status solidi (a)}\ }\textbf {\bibinfo {volume} {176}},\ \bibinfo
  {pages} {113} (\bibinfo {year} {1999})}\BibitemShut {NoStop}%
\bibitem [{\citenamefont {Tanaka}\ \emph {et~al.}(2003)\citenamefont {Tanaka},
  \citenamefont {Morishima}, \citenamefont {Bang}, \citenamefont {Ahn},
  \citenamefont {Sekiguchi},\ and\ \citenamefont {Akimoto}}]{Tanaka2003PSS}%
  \BibitemOpen
  \bibfield  {author} {\bibinfo {author} {\bibfnamefont {M.}~\bibnamefont
  {Tanaka}}, \bibinfo {author} {\bibfnamefont {S.}~\bibnamefont {Morishima}},
  \bibinfo {author} {\bibfnamefont {H.}~\bibnamefont {Bang}}, \bibinfo {author}
  {\bibfnamefont {J.~S.}\ \bibnamefont {Ahn}}, \bibinfo {author} {\bibfnamefont
  {T.}~\bibnamefont {Sekiguchi}},\ and\ \bibinfo {author} {\bibfnamefont
  {K.}~\bibnamefont {Akimoto}},\ }\bibfield  {title} {\bibinfo {title}
  {Low-energy charge-transfer state and optical properties of {Eu$^{3+}$}-doped
  {GaN}},\ }\href {https://doi.org/10.1002/pssc.200303447} {\bibfield
  {journal} {\bibinfo  {journal} {phys. status solidi (c)}\ }\textbf {\bibinfo
  {volume} {0}},\ \bibinfo {pages} {2639} (\bibinfo {year} {2003})}\BibitemShut
  {NoStop}%
\bibitem [{\citenamefont {Nyein}\ \emph {et~al.}(2003)\citenamefont {Nyein},
  \citenamefont {H{\"o}mmerich}, \citenamefont {Heikenfeld}, \citenamefont
  {Lee}, \citenamefont {Steckl},\ and\ \citenamefont {Zavada}}]{Nyein2003APL}%
  \BibitemOpen
  \bibfield  {author} {\bibinfo {author} {\bibfnamefont {E.~E.}\ \bibnamefont
  {Nyein}}, \bibinfo {author} {\bibfnamefont {U.}~\bibnamefont
  {H{\"o}mmerich}}, \bibinfo {author} {\bibfnamefont {J.}~\bibnamefont
  {Heikenfeld}}, \bibinfo {author} {\bibfnamefont {D.~S.}\ \bibnamefont {Lee}},
  \bibinfo {author} {\bibfnamefont {A.~J.}\ \bibnamefont {Steckl}},\ and\
  \bibinfo {author} {\bibfnamefont {J.~M.}\ \bibnamefont {Zavada}},\ }\bibfield
   {title} {\bibinfo {title} {{Spectral and time-resolved photoluminescence
  studies of Eu-doped GaN}},\ }\href {https://doi.org/10.1063/1.1560557}
  {\bibfield  {journal} {\bibinfo  {journal} {Appl. Phys. Lett.}\ }\textbf
  {\bibinfo {volume} {82}},\ \bibinfo {pages} {1655} (\bibinfo {year}
  {2003})}\BibitemShut {NoStop}%
\bibitem [{\citenamefont {Higuchi}\ \emph {et~al.}(2010)\citenamefont
  {Higuchi}, \citenamefont {Ishizumi}, \citenamefont {Sawahata}, \citenamefont
  {Akimoto},\ and\ \citenamefont {Kanemitsu}}]{Higuchi2010PRB}%
  \BibitemOpen
  \bibfield  {author} {\bibinfo {author} {\bibfnamefont {S.}~\bibnamefont
  {Higuchi}}, \bibinfo {author} {\bibfnamefont {A.}~\bibnamefont {Ishizumi}},
  \bibinfo {author} {\bibfnamefont {J.}~\bibnamefont {Sawahata}}, \bibinfo
  {author} {\bibfnamefont {K.}~\bibnamefont {Akimoto}},\ and\ \bibinfo {author}
  {\bibfnamefont {Y.}~\bibnamefont {Kanemitsu}},\ }\bibfield  {title} {\bibinfo
  {title} {Luminescence and energy-transfer mechanisms in {Eu}$^{3+}$-doped
  {GaN} epitaxial films},\ }\href {https://doi.org/10.1103/PhysRevB.81.035207}
  {\bibfield  {journal} {\bibinfo  {journal} {Phys. Rev. B}\ }\textbf {\bibinfo
  {volume} {81}},\ \bibinfo {pages} {035207} (\bibinfo {year}
  {2010})}\BibitemShut {NoStop}%
\bibitem [{\citenamefont {Li}\ \emph {et~al.}(2002)\citenamefont {Li},
  \citenamefont {Bang}, \citenamefont {Piao}, \citenamefont {Sawahata},\ and\
  \citenamefont {Akimoto}}]{Li2002JCG}%
  \BibitemOpen
  \bibfield  {author} {\bibinfo {author} {\bibfnamefont {Z.}~\bibnamefont
  {Li}}, \bibinfo {author} {\bibfnamefont {H.}~\bibnamefont {Bang}}, \bibinfo
  {author} {\bibfnamefont {G.}~\bibnamefont {Piao}}, \bibinfo {author}
  {\bibfnamefont {J.}~\bibnamefont {Sawahata}},\ and\ \bibinfo {author}
  {\bibfnamefont {K.}~\bibnamefont {Akimoto}},\ }\bibfield  {title} {\bibinfo
  {title} {{Growth of Eu-doped GaN by gas source molecular beam epitaxy and its
  optical properties}},\ }\href {https://doi.org/10.1016/S0022-0248(02)00952-1}
  {\bibfield  {journal} {\bibinfo  {journal} {J. Cryst. Growth}\ }\textbf
  {\bibinfo {volume} {240}},\ \bibinfo {pages} {382 } (\bibinfo {year}
  {2002})}\BibitemShut {NoStop}%
\bibitem [{\citenamefont {Sawahata}\ \emph {et~al.}(2005)\citenamefont
  {Sawahata}, \citenamefont {Bang}, \citenamefont {Seo},\ and\ \citenamefont
  {Akimoto}}]{Sawahata2005STAM}%
  \BibitemOpen
  \bibfield  {author} {\bibinfo {author} {\bibfnamefont {J.}~\bibnamefont
  {Sawahata}}, \bibinfo {author} {\bibfnamefont {H.}~\bibnamefont {Bang}},
  \bibinfo {author} {\bibfnamefont {J.}~\bibnamefont {Seo}},\ and\ \bibinfo
  {author} {\bibfnamefont {K.}~\bibnamefont {Akimoto}},\ }\bibfield  {title}
  {\bibinfo {title} {{Optical processes of red emission from Eu doped GaN}},\
  }\href {https://doi.org/10.1016/j.stam.2005.07.001} {\bibfield  {journal}
  {\bibinfo  {journal} {Sci. Technol. Adv. Mater.}\ }\textbf {\bibinfo {volume}
  {6}},\ \bibinfo {pages} {644} (\bibinfo {year} {2005})}\BibitemShut {NoStop}%
\bibitem [{\citenamefont {Dorenbos}\ and\ \citenamefont {van~der
  Kolk}(2006)}]{Dorenbos2006APL}%
  \BibitemOpen
  \bibfield  {author} {\bibinfo {author} {\bibfnamefont {P.}~\bibnamefont
  {Dorenbos}}\ and\ \bibinfo {author} {\bibfnamefont {E.}~\bibnamefont {van~der
  Kolk}},\ }\bibfield  {title} {\bibinfo {title} {{Location of lanthanide
  impurity levels in the III-V semiconductor GaN}},\ }\href
  {https://doi.org/10.1063/1.2336716} {\bibfield  {journal} {\bibinfo
  {journal} {Appl. Phys. Lett.}\ }\textbf {\bibinfo {volume} {89}},\ \bibinfo
  {pages} {061122} (\bibinfo {year} {2006})}\BibitemShut {NoStop}%
\bibitem [{\citenamefont {Ceperley}\ and\ \citenamefont
  {Alder}(1980)}]{LDA1980}%
  \BibitemOpen
  \bibfield  {author} {\bibinfo {author} {\bibfnamefont {D.~M.}\ \bibnamefont
  {Ceperley}}\ and\ \bibinfo {author} {\bibfnamefont {B.~J.}\ \bibnamefont
  {Alder}},\ }\bibfield  {title} {\bibinfo {title} {{Ground State of the
  Electron Gas by a Stochastic Method}},\ }\href
  {https://doi.org/10.1103/PhysRevLett.45.566} {\bibfield  {journal} {\bibinfo
  {journal} {Phys. Rev. Lett.}\ }\textbf {\bibinfo {volume} {45}},\ \bibinfo
  {pages} {566} (\bibinfo {year} {1980})}\BibitemShut {NoStop}%
\bibitem [{\citenamefont {Perdew}\ \emph {et~al.}(1992)\citenamefont {Perdew},
  \citenamefont {Chevary}, \citenamefont {Vosko}, \citenamefont {Jackson},
  \citenamefont {Pederson}, \citenamefont {Singh},\ and\ \citenamefont
  {Fiolhais}}]{PW91}%
  \BibitemOpen
  \bibfield  {author} {\bibinfo {author} {\bibfnamefont {J.~P.}\ \bibnamefont
  {Perdew}}, \bibinfo {author} {\bibfnamefont {J.~A.}\ \bibnamefont {Chevary}},
  \bibinfo {author} {\bibfnamefont {S.~H.}\ \bibnamefont {Vosko}}, \bibinfo
  {author} {\bibfnamefont {K.~A.}\ \bibnamefont {Jackson}}, \bibinfo {author}
  {\bibfnamefont {M.~R.}\ \bibnamefont {Pederson}}, \bibinfo {author}
  {\bibfnamefont {D.~J.}\ \bibnamefont {Singh}},\ and\ \bibinfo {author}
  {\bibfnamefont {C.}~\bibnamefont {Fiolhais}},\ }\bibfield  {title} {\bibinfo
  {title} {{Atoms, molecules, solids, and surfaces: Applications of the
  generalized gradient approximation for exchange and correlation}},\ }\href
  {https://doi.org/10.1103/PhysRevB.46.6671} {\bibfield  {journal} {\bibinfo
  {journal} {Phys. Rev. B}\ }\textbf {\bibinfo {volume} {46}},\ \bibinfo
  {pages} {6671} (\bibinfo {year} {1992})}\BibitemShut {NoStop}%
\bibitem [{\citenamefont {Anisimov}\ \emph {et~al.}(1991)\citenamefont
  {Anisimov}, \citenamefont {Zaanen},\ and\ \citenamefont
  {Andersen}}]{anisimov1991}%
  \BibitemOpen
  \bibfield  {author} {\bibinfo {author} {\bibfnamefont {V.~I.}\ \bibnamefont
  {Anisimov}}, \bibinfo {author} {\bibfnamefont {J.}~\bibnamefont {Zaanen}},\
  and\ \bibinfo {author} {\bibfnamefont {O.~K.}\ \bibnamefont {Andersen}},\
  }\bibfield  {title} {\bibinfo {title} {Hubbard-corrected density-functional
  theory},\ }\href {https://doi.org/10.1103/PhysRevB.44.943} {\bibfield
  {journal} {\bibinfo  {journal} {Phys. Rev. B}\ }\textbf {\bibinfo {volume}
  {44}},\ \bibinfo {pages} {943} (\bibinfo {year} {1991})}\BibitemShut
  {NoStop}%
\bibitem [{\citenamefont {Hoang}(2021)}]{Hoang2021PRM}%
  \BibitemOpen
  \bibfield  {author} {\bibinfo {author} {\bibfnamefont {K.}~\bibnamefont
  {Hoang}},\ }\bibfield  {title} {\bibinfo {title} {Tuning the valence and
  concentration of europium and luminescence centers in {GaN} through co-doping
  and defect association},\ }\href
  {https://doi.org/10.1103/PhysRevMaterials.5.034601} {\bibfield  {journal}
  {\bibinfo  {journal} {Phys. Rev. Materials}\ }\textbf {\bibinfo {volume}
  {5}},\ \bibinfo {pages} {034601} (\bibinfo {year} {2021})}\BibitemShut
  {NoStop}%
\bibitem [{\citenamefont {Heyd}\ \emph {et~al.}(2003)\citenamefont {Heyd},
  \citenamefont {Scuseria},\ and\ \citenamefont {Ernzerhof}}]{heyd:8207}%
  \BibitemOpen
  \bibfield  {author} {\bibinfo {author} {\bibfnamefont {J.}~\bibnamefont
  {Heyd}}, \bibinfo {author} {\bibfnamefont {G.~E.}\ \bibnamefont {Scuseria}},\
  and\ \bibinfo {author} {\bibfnamefont {M.}~\bibnamefont {Ernzerhof}},\
  }\bibfield  {title} {\bibinfo {title} {{Hybrid functionals based on a
  screened Coulomb potential}},\ }\href {https://doi.org/10.1063/1.1564060}
  {\bibfield  {journal} {\bibinfo  {journal} {J. Chem. Phys.}\ }\textbf
  {\bibinfo {volume} {118}},\ \bibinfo {pages} {8207} (\bibinfo {year}
  {2003})}\BibitemShut {NoStop}%
\bibitem [{\citenamefont {Hoang}(2015)}]{Hoang2015RRL}%
  \BibitemOpen
  \bibfield  {author} {\bibinfo {author} {\bibfnamefont {K.}~\bibnamefont
  {Hoang}},\ }\bibfield  {title} {\bibinfo {title} {{Hybrid density functional
  study of optically active Er$^{3+}$ centers in GaN}},\ }\href
  {https://doi.org/10.1002/pssr.201510269} {\bibfield  {journal} {\bibinfo
  {journal} {Phys. Status Solidi RRL}\ }\textbf {\bibinfo {volume} {9}},\
  \bibinfo {pages} {722} (\bibinfo {year} {2015})}\BibitemShut {NoStop}%
\bibitem [{\citenamefont {Hoang}(2016)}]{Hoang2016RRL}%
  \BibitemOpen
  \bibfield  {author} {\bibinfo {author} {\bibfnamefont {K.}~\bibnamefont
  {Hoang}},\ }\bibfield  {title} {\bibinfo {title} {{First-principles
  identification of defect levels in Er-doped GaN}},\ }\href
  {https://doi.org/10.1002/pssr.201600273} {\bibfield  {journal} {\bibinfo
  {journal} {Phys. Status Solidi RRL}\ }\textbf {\bibinfo {volume} {10}},\
  \bibinfo {pages} {915} (\bibinfo {year} {2016})}\BibitemShut {NoStop}%
\bibitem [{\citenamefont {Da~Silva}\ \emph {et~al.}(2007)\citenamefont
  {Da~Silva}, \citenamefont {Ganduglia-Pirovano}, \citenamefont {Sauer},
  \citenamefont {Bayer},\ and\ \citenamefont {Kresse}}]{DaSilva2007PRB}%
  \BibitemOpen
  \bibfield  {author} {\bibinfo {author} {\bibfnamefont {J.~L.~F.}\
  \bibnamefont {Da~Silva}}, \bibinfo {author} {\bibfnamefont {M.~V.}\
  \bibnamefont {Ganduglia-Pirovano}}, \bibinfo {author} {\bibfnamefont
  {J.}~\bibnamefont {Sauer}}, \bibinfo {author} {\bibfnamefont
  {V.}~\bibnamefont {Bayer}},\ and\ \bibinfo {author} {\bibfnamefont
  {G.}~\bibnamefont {Kresse}},\ }\bibfield  {title} {\bibinfo {title} {Hybrid
  functionals applied to rare-earth oxides: {T}he example of ceria},\ }\href
  {https://doi.org/10.1103/PhysRevB.75.045121} {\bibfield  {journal} {\bibinfo
  {journal} {Phys. Rev. B}\ }\textbf {\bibinfo {volume} {75}},\ \bibinfo
  {pages} {045121} (\bibinfo {year} {2007})}\BibitemShut {NoStop}%
\bibitem [{\citenamefont {Freysoldt}\ \emph {et~al.}(2014)\citenamefont
  {Freysoldt}, \citenamefont {Grabowski}, \citenamefont {Hickel}, \citenamefont
  {Neugebauer}, \citenamefont {Kresse}, \citenamefont {Janotti},\ and\
  \citenamefont {{Van de Walle}}}]{Freysoldt2014RMP}%
  \BibitemOpen
  \bibfield  {author} {\bibinfo {author} {\bibfnamefont {C.}~\bibnamefont
  {Freysoldt}}, \bibinfo {author} {\bibfnamefont {B.}~\bibnamefont
  {Grabowski}}, \bibinfo {author} {\bibfnamefont {T.}~\bibnamefont {Hickel}},
  \bibinfo {author} {\bibfnamefont {J.}~\bibnamefont {Neugebauer}}, \bibinfo
  {author} {\bibfnamefont {G.}~\bibnamefont {Kresse}}, \bibinfo {author}
  {\bibfnamefont {A.}~\bibnamefont {Janotti}},\ and\ \bibinfo {author}
  {\bibfnamefont {C.~G.}\ \bibnamefont {{Van de Walle}}},\ }\bibfield  {title}
  {\bibinfo {title} {First-principles calculations for point defects in
  solids},\ }\href {https://doi.org/10.1103/RevModPhys.86.253} {\bibfield
  {journal} {\bibinfo  {journal} {Rev. Mod. Phys.}\ }\textbf {\bibinfo {volume}
  {86}},\ \bibinfo {pages} {253} (\bibinfo {year} {2014})}\BibitemShut
  {NoStop}%
\bibitem [{\citenamefont {Freysoldt}\ \emph {et~al.}(2009)\citenamefont
  {Freysoldt}, \citenamefont {Neugebauer},\ and\ \citenamefont {{Van de
  Walle}}}]{Freysoldt}%
  \BibitemOpen
  \bibfield  {author} {\bibinfo {author} {\bibfnamefont {C.}~\bibnamefont
  {Freysoldt}}, \bibinfo {author} {\bibfnamefont {J.}~\bibnamefont
  {Neugebauer}},\ and\ \bibinfo {author} {\bibfnamefont {C.~G.}\ \bibnamefont
  {{Van de Walle}}},\ }\bibfield  {title} {\bibinfo {title} {{Fully \textit{Ab
  Initio} Finite-Size Corrections for Charged-Defect Supercell Calculations}},\
  }\href {https://doi.org/10.1103/PhysRevLett.102.016402} {\bibfield  {journal}
  {\bibinfo  {journal} {Phys. Rev. Lett.}\ }\textbf {\bibinfo {volume} {102}},\
  \bibinfo {pages} {016402} (\bibinfo {year} {2009})}\BibitemShut {NoStop}%
\bibitem [{\citenamefont {Freysoldt}\ \emph {et~al.}(2011)\citenamefont
  {Freysoldt}, \citenamefont {Neugebauer},\ and\ \citenamefont {{Van de
  Walle}}}]{Freysoldt11}%
  \BibitemOpen
  \bibfield  {author} {\bibinfo {author} {\bibfnamefont {C.}~\bibnamefont
  {Freysoldt}}, \bibinfo {author} {\bibfnamefont {J.}~\bibnamefont
  {Neugebauer}},\ and\ \bibinfo {author} {\bibfnamefont {C.~G.}\ \bibnamefont
  {{Van de Walle}}},\ }\bibfield  {title} {\bibinfo {title} {Electrostatic
  interactions between charged defects in supercells},\ }\href
  {https://doi.org/10.1002/pssb.201046289} {\bibfield  {journal} {\bibinfo
  {journal} {phys. status solidi (b)}\ }\textbf {\bibinfo {volume} {248}},\
  \bibinfo {pages} {1067} (\bibinfo {year} {2011})}\BibitemShut {NoStop}%
\bibitem [{\citenamefont {Kresse}\ and\ \citenamefont {Joubert}(1999)}]{PAW2}%
  \BibitemOpen
  \bibfield  {author} {\bibinfo {author} {\bibfnamefont {G.}~\bibnamefont
  {Kresse}}\ and\ \bibinfo {author} {\bibfnamefont {D.}~\bibnamefont
  {Joubert}},\ }\bibfield  {title} {\bibinfo {title} {From ultrasoft
  pseudopotentials to the projector augmented-wave method},\ }\href
  {https://doi.org/10.1103/PhysRevB.59.1758} {\bibfield  {journal} {\bibinfo
  {journal} {Phys. Rev. B}\ }\textbf {\bibinfo {volume} {59}},\ \bibinfo
  {pages} {1758} (\bibinfo {year} {1999})}\BibitemShut {NoStop}%
\bibitem [{\citenamefont {Kresse}\ and\ \citenamefont
  {Furthm\"uller}(1996)}]{VASP2}%
  \BibitemOpen
  \bibfield  {author} {\bibinfo {author} {\bibfnamefont {G.}~\bibnamefont
  {Kresse}}\ and\ \bibinfo {author} {\bibfnamefont {J.}~\bibnamefont
  {Furthm\"uller}},\ }\bibfield  {title} {\bibinfo {title} {Efficient iterative
  schemes for ab initio total-energy calculations using a plane-wave basis
  set},\ }\href {https://doi.org/10.1103/PhysRevB.54.11169} {\bibfield
  {journal} {\bibinfo  {journal} {Phys. Rev. B}\ }\textbf {\bibinfo {volume}
  {54}},\ \bibinfo {pages} {11169} (\bibinfo {year} {1996})}\BibitemShut
  {NoStop}%
\bibitem [{\citenamefont {Kordis}\ and\ \citenamefont
  {Gingerich}(1977)}]{Kordis1977JNM}%
  \BibitemOpen
  \bibfield  {author} {\bibinfo {author} {\bibfnamefont {J.}~\bibnamefont
  {Kordis}}\ and\ \bibinfo {author} {\bibfnamefont {K.~A.}\ \bibnamefont
  {Gingerich}},\ }\bibfield  {title} {\bibinfo {title} {Heats of vaporization
  and standard heats of formation of rare earth mononitrides},\ }\href
  {https://doi.org/10.1016/0022-3115(77)90147-7} {\bibfield  {journal}
  {\bibinfo  {journal} {J. Nucl. Mater.}\ }\textbf {\bibinfo {volume} {66}},\
  \bibinfo {pages} {197} (\bibinfo {year} {1977})}\BibitemShut {NoStop}%
\bibitem [{\citenamefont {Phani}\ \emph {et~al.}(1980)\citenamefont {Phani},
  \citenamefont {Lebowitz},\ and\ \citenamefont {Kalos}}]{Phani1980PRB}%
  \BibitemOpen
  \bibfield  {author} {\bibinfo {author} {\bibfnamefont {M.~K.}\ \bibnamefont
  {Phani}}, \bibinfo {author} {\bibfnamefont {J.~L.}\ \bibnamefont
  {Lebowitz}},\ and\ \bibinfo {author} {\bibfnamefont {M.~H.}\ \bibnamefont
  {Kalos}},\ }\bibfield  {title} {\bibinfo {title} {{Monte Carlo studies of an
  {\it fcc} Ising antiferromagnet with nearest- and next-nearest-neighbor
  interactions}},\ }\href {https://doi.org/10.1103/PhysRevB.21.4027} {\bibfield
   {journal} {\bibinfo  {journal} {Phys. Rev. B}\ }\textbf {\bibinfo {volume}
  {21}},\ \bibinfo {pages} {4027} (\bibinfo {year} {1980})}\BibitemShut
  {NoStop}%
\bibitem [{\citenamefont {Hoang}\ \emph {et~al.}(2007)\citenamefont {Hoang},
  \citenamefont {Mahanti}, \citenamefont {Salvador},\ and\ \citenamefont
  {Kanatzidis}}]{Hoang2007PRL}%
  \BibitemOpen
  \bibfield  {author} {\bibinfo {author} {\bibfnamefont {K.}~\bibnamefont
  {Hoang}}, \bibinfo {author} {\bibfnamefont {S.~D.}\ \bibnamefont {Mahanti}},
  \bibinfo {author} {\bibfnamefont {J.~R.}\ \bibnamefont {Salvador}},\ and\
  \bibinfo {author} {\bibfnamefont {M.~G.}\ \bibnamefont {Kanatzidis}},\
  }\bibfield  {title} {\bibinfo {title} {{Atomic Ordering and Gap Formation in
  Ag-Sb-Based Ternary Chalcogenides}},\ }\href
  {https://doi.org/10.1103/PhysRevLett.99.156403} {\bibfield  {journal}
  {\bibinfo  {journal} {Phys. Rev. Lett.}\ }\textbf {\bibinfo {volume} {99}},\
  \bibinfo {pages} {156403} (\bibinfo {year} {2007})}\BibitemShut {NoStop}%
\bibitem [{\citenamefont {Schulz}\ and\ \citenamefont
  {Thiemann}(1977)}]{Schulz1977SSC}%
  \BibitemOpen
  \bibfield  {author} {\bibinfo {author} {\bibfnamefont {H.}~\bibnamefont
  {Schulz}}\ and\ \bibinfo {author} {\bibfnamefont {K.~H.}\ \bibnamefont
  {Thiemann}},\ }\bibfield  {title} {\bibinfo {title} {{Crystal structure
  refinement of AlN and GaN}},\ }\href
  {https://doi.org/10.1016/0038-1098(77)90959-0} {\bibfield  {journal}
  {\bibinfo  {journal} {Solid State Commun.}\ }\textbf {\bibinfo {volume}
  {23}},\ \bibinfo {pages} {815} (\bibinfo {year} {1977})}\BibitemShut
  {NoStop}%
\bibitem [{\citenamefont {Lyons}\ \emph {et~al.}(2021)\citenamefont {Lyons},
  \citenamefont {Wickramaratne},\ and\ \citenamefont {Van~de
  Walle}}]{Lyons2021JAP}%
  \BibitemOpen
  \bibfield  {author} {\bibinfo {author} {\bibfnamefont {J.~L.}\ \bibnamefont
  {Lyons}}, \bibinfo {author} {\bibfnamefont {D.}~\bibnamefont
  {Wickramaratne}},\ and\ \bibinfo {author} {\bibfnamefont {C.~G.}\
  \bibnamefont {Van~de Walle}},\ }\bibfield  {title} {\bibinfo {title} {A
  first-principles understanding of point defects and impurities in {GaN}},\
  }\href {https://doi.org/10.1063/5.0041506} {\bibfield  {journal} {\bibinfo
  {journal} {J. Appl. Phys.}\ }\textbf {\bibinfo {volume} {129}},\ \bibinfo
  {pages} {111101} (\bibinfo {year} {2021})}\BibitemShut {NoStop}%
\bibitem [{\citenamefont {Svane}\ \emph {et~al.}(2006)\citenamefont {Svane},
  \citenamefont {Christensen}, \citenamefont {Petit}, \citenamefont {Szotek},\
  and\ \citenamefont {Temmerman}}]{Svane2006}%
  \BibitemOpen
  \bibfield  {author} {\bibinfo {author} {\bibfnamefont {A.}~\bibnamefont
  {Svane}}, \bibinfo {author} {\bibfnamefont {N.~E.}\ \bibnamefont
  {Christensen}}, \bibinfo {author} {\bibfnamefont {L.}~\bibnamefont {Petit}},
  \bibinfo {author} {\bibfnamefont {Z.}~\bibnamefont {Szotek}},\ and\ \bibinfo
  {author} {\bibfnamefont {W.~M.}\ \bibnamefont {Temmerman}},\ }\bibfield
  {title} {\bibinfo {title} {{Electronic structure of rare-earth impurities in
  GaAs and GaN}},\ }\href {https://doi.org/10.1103/PhysRevB.74.165204}
  {\bibfield  {journal} {\bibinfo  {journal} {Phys. Rev. B}\ }\textbf {\bibinfo
  {volume} {74}},\ \bibinfo {pages} {165204} (\bibinfo {year}
  {2006})}\BibitemShut {NoStop}%
\bibitem [{\citenamefont {Sanna}\ \emph {et~al.}(2009)\citenamefont {Sanna},
  \citenamefont {Schmidt}, \citenamefont {Frauenheim},\ and\ \citenamefont
  {Gerstmann}}]{Sanna2009}%
  \BibitemOpen
  \bibfield  {author} {\bibinfo {author} {\bibfnamefont {S.}~\bibnamefont
  {Sanna}}, \bibinfo {author} {\bibfnamefont {W.~G.}\ \bibnamefont {Schmidt}},
  \bibinfo {author} {\bibfnamefont {T.}~\bibnamefont {Frauenheim}},\ and\
  \bibinfo {author} {\bibfnamefont {U.}~\bibnamefont {Gerstmann}},\ }\bibfield
  {title} {\bibinfo {title} {{Rare-earth defect pairs in GaN: LDA$+U$
  calculations}},\ }\href {https://doi.org/10.1103/PhysRevB.80.104120}
  {\bibfield  {journal} {\bibinfo  {journal} {Phys. Rev. B}\ }\textbf {\bibinfo
  {volume} {80}},\ \bibinfo {pages} {104120} (\bibinfo {year}
  {2009})}\BibitemShut {NoStop}%
\bibitem [{\citenamefont {Mitchell}\ \emph {et~al.}(2017)\citenamefont
  {Mitchell}, \citenamefont {Koizumi}, \citenamefont {Nunokawa}, \citenamefont
  {Wakamatsu}, \citenamefont {Lee}, \citenamefont {Saitoh}, \citenamefont
  {Timmerman}, \citenamefont {Kuboshima}, \citenamefont {Mogi}, \citenamefont
  {Higashi}, \citenamefont {Kikukawa}, \citenamefont {Ofuchi}, \citenamefont
  {Honma},\ and\ \citenamefont {Fujiwara}}]{Mitchell2017MCP}%
  \BibitemOpen
  \bibfield  {author} {\bibinfo {author} {\bibfnamefont {B.}~\bibnamefont
  {Mitchell}}, \bibinfo {author} {\bibfnamefont {A.}~\bibnamefont {Koizumi}},
  \bibinfo {author} {\bibfnamefont {T.}~\bibnamefont {Nunokawa}}, \bibinfo
  {author} {\bibfnamefont {R.}~\bibnamefont {Wakamatsu}}, \bibinfo {author}
  {\bibfnamefont {D.}~\bibnamefont {Lee}}, \bibinfo {author} {\bibfnamefont
  {Y.}~\bibnamefont {Saitoh}}, \bibinfo {author} {\bibfnamefont
  {D.}~\bibnamefont {Timmerman}}, \bibinfo {author} {\bibfnamefont
  {Y.}~\bibnamefont {Kuboshima}}, \bibinfo {author} {\bibfnamefont
  {T.}~\bibnamefont {Mogi}}, \bibinfo {author} {\bibfnamefont {S.}~\bibnamefont
  {Higashi}}, \bibinfo {author} {\bibfnamefont {K.}~\bibnamefont {Kikukawa}},
  \bibinfo {author} {\bibfnamefont {H.}~\bibnamefont {Ofuchi}}, \bibinfo
  {author} {\bibfnamefont {T.}~\bibnamefont {Honma}},\ and\ \bibinfo {author}
  {\bibfnamefont {Y.}~\bibnamefont {Fujiwara}},\ }\bibfield  {title} {\bibinfo
  {title} {{Synthesis and characterization of a liquid Eu precursor (EuCp$^{\rm
  pm}_2$) allowing for valence control of Eu ions doped into GaN by
  organometallic vapor phase epitaxy}},\ }\href
  {https://doi.org/10.1016/j.matchemphys.2017.02.021} {\bibfield  {journal}
  {\bibinfo  {journal} {Mater. Chem. Phys.}\ }\textbf {\bibinfo {volume}
  {193}},\ \bibinfo {pages} {140} (\bibinfo {year} {2017})}\BibitemShut
  {NoStop}%
\bibitem [{\citenamefont {Nunokawa}\ \emph {et~al.}(2020)\citenamefont
  {Nunokawa}, \citenamefont {Fujiwara}, \citenamefont {Miyata}, \citenamefont
  {Fujimura}, \citenamefont {Sakurai}, \citenamefont {Ohta}, \citenamefont
  {Masago}, \citenamefont {Shinya}, \citenamefont {Fukushima}, \citenamefont
  {Sato},\ and\ \citenamefont {{Katayama-Yoshida}}}]{Nunokawa2020JAP}%
  \BibitemOpen
  \bibfield  {author} {\bibinfo {author} {\bibfnamefont {T.}~\bibnamefont
  {Nunokawa}}, \bibinfo {author} {\bibfnamefont {Y.}~\bibnamefont {Fujiwara}},
  \bibinfo {author} {\bibfnamefont {Y.}~\bibnamefont {Miyata}}, \bibinfo
  {author} {\bibfnamefont {N.}~\bibnamefont {Fujimura}}, \bibinfo {author}
  {\bibfnamefont {T.}~\bibnamefont {Sakurai}}, \bibinfo {author} {\bibfnamefont
  {H.}~\bibnamefont {Ohta}}, \bibinfo {author} {\bibfnamefont {A.}~\bibnamefont
  {Masago}}, \bibinfo {author} {\bibfnamefont {H.}~\bibnamefont {Shinya}},
  \bibinfo {author} {\bibfnamefont {T.}~\bibnamefont {Fukushima}}, \bibinfo
  {author} {\bibfnamefont {K.}~\bibnamefont {Sato}},\ and\ \bibinfo {author}
  {\bibfnamefont {H.}~\bibnamefont {{Katayama-Yoshida}}},\ }\bibfield  {title}
  {\bibinfo {title} {{Valence states and the magnetism of Eu ions in Eu-doped
  GaN}},\ }\href {https://doi.org/10.1063/1.5135743} {\bibfield  {journal}
  {\bibinfo  {journal} {J. Appl. Phys.}\ }\textbf {\bibinfo {volume} {127}},\
  \bibinfo {pages} {083901} (\bibinfo {year} {2020})}\BibitemShut {NoStop}%
\bibitem [{\citenamefont {Hoang}\ and\ \citenamefont
  {Johannes}(2018)}]{Hoang2018JPCM}%
  \BibitemOpen
  \bibfield  {author} {\bibinfo {author} {\bibfnamefont {K.}~\bibnamefont
  {Hoang}}\ and\ \bibinfo {author} {\bibfnamefont {M.~D.}\ \bibnamefont
  {Johannes}},\ }\bibfield  {title} {\bibinfo {title} {Defect physics in
  complex energy materials},\ }\href {https://doi.org/10.1088/1361-648x/aacb05}
  {\bibfield  {journal} {\bibinfo  {journal} {J. Phys.: Condens. Matter}\
  }\textbf {\bibinfo {volume} {30}},\ \bibinfo {pages} {293001} (\bibinfo
  {year} {2018})}\BibitemShut {NoStop}%
\bibitem [{\citenamefont {Alkauskas}\ \emph {et~al.}(2016)\citenamefont
  {Alkauskas}, \citenamefont {McCluskey},\ and\ \citenamefont {Van~de
  Walle}}]{Alkauskas2016JAP}%
  \BibitemOpen
  \bibfield  {author} {\bibinfo {author} {\bibfnamefont {A.}~\bibnamefont
  {Alkauskas}}, \bibinfo {author} {\bibfnamefont {M.~D.}\ \bibnamefont
  {McCluskey}},\ and\ \bibinfo {author} {\bibfnamefont {C.~G.}\ \bibnamefont
  {Van~de Walle}},\ }\bibfield  {title} {\bibinfo {title} {{Defects in
  semiconductors--Combining experiment and theory}},\ }\href
  {https://doi.org/10.1063/1.4948245} {\bibfield  {journal} {\bibinfo
  {journal} {J. Appl. Phys.}\ }\textbf {\bibinfo {volume} {119}},\ \bibinfo
  {pages} {181101} (\bibinfo {year} {2016})}\BibitemShut {NoStop}%
\bibitem [{\citenamefont {Dorenbos}(2017)}]{Dorenbos2017OM}%
  \BibitemOpen
  \bibfield  {author} {\bibinfo {author} {\bibfnamefont {P.}~\bibnamefont
  {Dorenbos}},\ }\bibfield  {title} {\bibinfo {title} {Charge transfer bands in
  optical materials and related defect level location},\ }\href
  {https://doi.org/10.1016/j.optmat.2017.03.061} {\bibfield  {journal}
  {\bibinfo  {journal} {Opt. Mater.}\ }\textbf {\bibinfo {volume} {69}},\
  \bibinfo {pages} {8} (\bibinfo {year} {2017})}\BibitemShut {NoStop}%
\bibitem [{\citenamefont {Wickramaratne}\ \emph {et~al.}(2019)\citenamefont
  {Wickramaratne}, \citenamefont {Shen}, \citenamefont {Dreyer}, \citenamefont
  {Alkauskas},\ and\ \citenamefont {Van~de Walle}}]{Wickramaratne2019PRB}%
  \BibitemOpen
  \bibfield  {author} {\bibinfo {author} {\bibfnamefont {D.}~\bibnamefont
  {Wickramaratne}}, \bibinfo {author} {\bibfnamefont {J.-X.}\ \bibnamefont
  {Shen}}, \bibinfo {author} {\bibfnamefont {C.~E.}\ \bibnamefont {Dreyer}},
  \bibinfo {author} {\bibfnamefont {A.}~\bibnamefont {Alkauskas}},\ and\
  \bibinfo {author} {\bibfnamefont {C.~G.}\ \bibnamefont {Van~de Walle}},\
  }\bibfield  {title} {\bibinfo {title} {{Electrical and optical properties of
  iron in GaN, AlN, and InN}},\ }\href
  {https://doi.org/10.1103/PhysRevB.99.205202} {\bibfield  {journal} {\bibinfo
  {journal} {Phys. Rev. B}\ }\textbf {\bibinfo {volume} {99}},\ \bibinfo
  {pages} {205202} (\bibinfo {year} {2019})}\BibitemShut {NoStop}%
\bibitem [{\citenamefont {Huang}\ and\ \citenamefont {Rhys}(1950)}]{Huang1950}%
  \BibitemOpen
  \bibfield  {author} {\bibinfo {author} {\bibfnamefont {K.}~\bibnamefont
  {Huang}}\ and\ \bibinfo {author} {\bibfnamefont {A.}~\bibnamefont {Rhys}},\
  }\bibfield  {title} {\bibinfo {title} {Theory of light absorption and
  non-radiative transitions in {$F$}-centres},\ }\href
  {https://doi.org/10.1098/rspa.1950.0184} {\bibfield  {journal} {\bibinfo
  {journal} {Proc. R. Soc. Lond. A}\ }\textbf {\bibinfo {volume} {204}},\
  \bibinfo {pages} {406} (\bibinfo {year} {1950})}\BibitemShut {NoStop}%
\bibitem [{\citenamefont {Alkauskas}\ \emph {et~al.}(2012)\citenamefont
  {Alkauskas}, \citenamefont {Lyons}, \citenamefont {Steiauf},\ and\
  \citenamefont {Van~de Walle}}]{Alkauskas2012PRL}%
  \BibitemOpen
  \bibfield  {author} {\bibinfo {author} {\bibfnamefont {A.}~\bibnamefont
  {Alkauskas}}, \bibinfo {author} {\bibfnamefont {J.~L.}\ \bibnamefont
  {Lyons}}, \bibinfo {author} {\bibfnamefont {D.}~\bibnamefont {Steiauf}},\
  and\ \bibinfo {author} {\bibfnamefont {C.~G.}\ \bibnamefont {Van~de Walle}},\
  }\bibfield  {title} {\bibinfo {title} {{First-Principles Calculations of
  Luminescence Spectrum Line Shapes for Defects in Semiconductors: The Example
  of GaN and ZnO}},\ }\href {https://doi.org/10.1103/PhysRevLett.109.267401}
  {\bibfield  {journal} {\bibinfo  {journal} {Phys. Rev. Lett.}\ }\textbf
  {\bibinfo {volume} {109}},\ \bibinfo {pages} {267401} (\bibinfo {year}
  {2012})}\BibitemShut {NoStop}%
\bibitem [{\citenamefont {{van Pieterson}}\ \emph {et~al.}(2000)\citenamefont
  {{van Pieterson}}, \citenamefont {Heeroma}, \citenamefont {{de Heer}},\ and\
  \citenamefont {Meijerink}}]{vanPieterson2000JL}%
  \BibitemOpen
  \bibfield  {author} {\bibinfo {author} {\bibfnamefont {L.}~\bibnamefont {{van
  Pieterson}}}, \bibinfo {author} {\bibfnamefont {M.}~\bibnamefont {Heeroma}},
  \bibinfo {author} {\bibfnamefont {E.}~\bibnamefont {{de Heer}}},\ and\
  \bibinfo {author} {\bibfnamefont {A.}~\bibnamefont {Meijerink}},\ }\bibfield
  {title} {\bibinfo {title} {Charge transfer luminescence of {Yb$^{3+}$}},\
  }\href {https://doi.org/10.1016/S0022-2313(00)00214-3} {\bibfield  {journal}
  {\bibinfo  {journal} {J. Lumin.}\ }\textbf {\bibinfo {volume} {91}},\
  \bibinfo {pages} {177} (\bibinfo {year} {2000})}\BibitemShut {NoStop}%
\bibitem [{\citenamefont {Nakazawa}(1978)}]{Nakazawa1978CPL}%
  \BibitemOpen
  \bibfield  {author} {\bibinfo {author} {\bibfnamefont {E.}~\bibnamefont
  {Nakazawa}},\ }\bibfield  {title} {\bibinfo {title} {{Charge-transfer type
  luminescence of Yb$^{3+}$ ions in LuPO$_4$ and YPO$_4$}},\ }\href
  {https://doi.org/10.1016/0009-2614(78)80210-3} {\bibfield  {journal}
  {\bibinfo  {journal} {Chem. Phys. Lett.}\ }\textbf {\bibinfo {volume} {56}},\
  \bibinfo {pages} {161} (\bibinfo {year} {1978})}\BibitemShut {NoStop}%
\bibitem [{\citenamefont {Nakazawa}(1979)}]{Nakazawa1979JL}%
  \BibitemOpen
  \bibfield  {author} {\bibinfo {author} {\bibfnamefont {E.}~\bibnamefont
  {Nakazawa}},\ }\bibfield  {title} {\bibinfo {title} {{Charge transfer type
  luminescence of Yb$^{3+}$ ions in RPO$_4$ and R$_2$O$_2$S (R=Y, La, and
  Lu)}},\ }\href {https://doi.org/10.1016/0022-2313(79)90119-4} {\bibfield
  {journal} {\bibinfo  {journal} {J. Lumin.}\ }\textbf {\bibinfo {volume}
  {18-19}},\ \bibinfo {pages} {272} (\bibinfo {year} {1979})}\BibitemShut
  {NoStop}%
\bibitem [{\citenamefont {Danielson}\ \emph {et~al.}(1998)\citenamefont
  {Danielson}, \citenamefont {Devenney}, \citenamefont {Giaquinta},
  \citenamefont {Golden}, \citenamefont {Haushalter}, \citenamefont
  {McFarland}, \citenamefont {Poojary}, \citenamefont {Reaves}, \citenamefont
  {Weinberg},\ and\ \citenamefont {Wu}}]{Danielson1998Science}%
  \BibitemOpen
  \bibfield  {author} {\bibinfo {author} {\bibfnamefont {E.}~\bibnamefont
  {Danielson}}, \bibinfo {author} {\bibfnamefont {M.}~\bibnamefont {Devenney}},
  \bibinfo {author} {\bibfnamefont {D.~M.}\ \bibnamefont {Giaquinta}}, \bibinfo
  {author} {\bibfnamefont {J.~H.}\ \bibnamefont {Golden}}, \bibinfo {author}
  {\bibfnamefont {R.~C.}\ \bibnamefont {Haushalter}}, \bibinfo {author}
  {\bibfnamefont {E.~W.}\ \bibnamefont {McFarland}}, \bibinfo {author}
  {\bibfnamefont {D.~M.}\ \bibnamefont {Poojary}}, \bibinfo {author}
  {\bibfnamefont {C.~M.}\ \bibnamefont {Reaves}}, \bibinfo {author}
  {\bibfnamefont {W.~H.}\ \bibnamefont {Weinberg}},\ and\ \bibinfo {author}
  {\bibfnamefont {X.~D.}\ \bibnamefont {Wu}},\ }\bibfield  {title} {\bibinfo
  {title} {{A Rare-Earth Phosphor Containing One-Dimensional Chains Identified
  Through Combinatorial Methods}},\ }\href
  {https://doi.org/10.1126/science.279.5352.837} {\bibfield  {journal}
  {\bibinfo  {journal} {Science}\ }\textbf {\bibinfo {volume} {279}},\ \bibinfo
  {pages} {837} (\bibinfo {year} {1998})}\BibitemShut {NoStop}%
\bibitem [{\citenamefont {Razinkovas}\ \emph {et~al.}(2021)\citenamefont
  {Razinkovas}, \citenamefont {Maciaszek}, \citenamefont {Reinhard},
  \citenamefont {Doherty},\ and\ \citenamefont
  {Alkauskas}}]{Razinkovas2021PRB}%
  \BibitemOpen
  \bibfield  {author} {\bibinfo {author} {\bibfnamefont {L.}~\bibnamefont
  {Razinkovas}}, \bibinfo {author} {\bibfnamefont {M.}~\bibnamefont
  {Maciaszek}}, \bibinfo {author} {\bibfnamefont {F.}~\bibnamefont {Reinhard}},
  \bibinfo {author} {\bibfnamefont {M.~W.}\ \bibnamefont {Doherty}},\ and\
  \bibinfo {author} {\bibfnamefont {A.}~\bibnamefont {Alkauskas}},\ }\bibfield
  {title} {\bibinfo {title} {{Photoionization of negatively charged NV centers
  in diamond: Theory and {\it ab initio} calculations}},\ }\href
  {https://doi.org/10.1103/PhysRevB.104.235301} {\bibfield  {journal} {\bibinfo
   {journal} {Phys. Rev. B}\ }\textbf {\bibinfo {volume} {104}},\ \bibinfo
  {pages} {235301} (\bibinfo {year} {2021})}\BibitemShut {NoStop}%
\bibitem [{\citenamefont {Dreyer}\ \emph {et~al.}(2020)\citenamefont {Dreyer},
  \citenamefont {Alkauskas}, \citenamefont {Lyons},\ and\ \citenamefont {Van~de
  Walle}}]{Dreyer2020PRB}%
  \BibitemOpen
  \bibfield  {author} {\bibinfo {author} {\bibfnamefont {C.~E.}\ \bibnamefont
  {Dreyer}}, \bibinfo {author} {\bibfnamefont {A.}~\bibnamefont {Alkauskas}},
  \bibinfo {author} {\bibfnamefont {J.~L.}\ \bibnamefont {Lyons}},\ and\
  \bibinfo {author} {\bibfnamefont {C.~G.}\ \bibnamefont {Van~de Walle}},\
  }\bibfield  {title} {\bibinfo {title} {Radiative capture rates at deep
  defects from electronic structure calculations},\ }\href
  {https://doi.org/10.1103/PhysRevB.102.085305} {\bibfield  {journal} {\bibinfo
   {journal} {Phys. Rev. B}\ }\textbf {\bibinfo {volume} {102}},\ \bibinfo
  {pages} {085305} (\bibinfo {year} {2020})}\BibitemShut {NoStop}%
\end{thebibliography}

%

\end{document}